\documentclass[notitlepage,english,aps,prx,tightenlines,floatfix,longbibliography,10pt,citeautoscript,twocolumn,colorlinks=true,urlcolor=blue,citecolor=blue, linkcolor=blue, superscriptaddress]{revtex4-2}

\usepackage{graphicx}
\usepackage{dcolumn}
\usepackage{bm}
\usepackage{color}
\usepackage[dvipsnames]{xcolor}
\usepackage{physics}
\usepackage{orcidlink}
\usepackage{hyperref}
\usepackage{amssymb,ulem,amsmath}
\usepackage{float}
\usepackage{natbib}
\usepackage{comment}
\usepackage[T1]{fontenc}
\usepackage{lmodern}     
\usepackage{cmap}        
\usepackage{fix-cm} 	 
\newcommand{\dg}{\dagger}
\newcommand{\ham}{\mathcal{\hat H}}

\newcommand{\sumj}[0]{\sum_{j=0}^{L-1}}

\newcommand{\up}{\uparrow}
\newcommand{\down}{\downarrow}
\newcommand{\dw}{\downarrow}

\newcommand{\vac}[0]{\ket{\emptyset}}

\newcommand{\ddagup}[1]{\hat d^\dg_{#1\up}}
\newcommand{\ddagdown}[1]{\hat d^\dg_{#1\down}}

\newcommand{\Bplus}[1]{\hat B^+_{#1}}

\newcommand{\Bijplus}[2]{\hat B^+_{\langle #1, #2 \rangle}}

\begin{document}

\title{Coexistence of ergodic and nonergodic behavior and level spacing statistics in a one-dimensional model of a flat band superconductor}

\author{Meri Teeriaho}
\affiliation{Department of Physics, University of Helsinki, Gustaf H\"allstr\"omin katu 2, FI-00014 Helsinki, Finland}
\affiliation{Department of Applied Physics, Aalto University, FI-00076 Aalto, Finland}

\author{Ville-Vertti Linho}
\affiliation{Department of Physics, University of Helsinki, Gustaf H\"allstr\"omin katu 2, FI-00014 Helsinki, Finland}
\affiliation{Department of Applied Physics, Aalto University, FI-00076 Aalto, Finland}

\author{Koushik Swaminathan\orcidlink{https://orcid.org/0000-0003-4932-9977}}
\affiliation{Department of Applied Physics, Aalto University, FI-00076 Aalto, Finland}

\author{Sebastiano Peotta\orcidlink{https://orcid.org/0000-0002-9947-1261}}
\email{sebastiano.peotta@aalto.fi}	
\affiliation{Department of Applied Physics, Aalto University, FI-00076 Aalto, Finland}

\begin{abstract}
    Motivated by recent studies of the projected dice lattice Hamiltonian [\href{https://journals.aps.org/prresearch/abstract/10.1103/PhysRevResearch.5.043215}{K. Swaminathan et al., Phys. Rev. Res. \textbf{5}, 043215 (2023)}], we introduce the on-site/bond singlet (OBS) model, a one-dimensional model of a flat band superconductor, in order to better understand the quasiparticle localization and interesting coexistence of ergodic and nonergodic behavior present in the former model. The OBS model is the sum of terms that have direct counterparts in the projected dice lattice Hamiltonian, each of which is parameterized by a coupling constant. Exact diagonalization reveals that the energy spectrum and nonequilibrium dynamics of the OBS model are essentially the same as that of the dice lattice for some values of the coupling constants. The quasiparticle localization and breaking of ergodicity manifest in a striking manner in the level spacing distribution. Its near Poissonian form provides evidence for the existence of local integrals of motion and establishes the OBS model as a  nontrivial integrable generalization of the projected Creutz ladder Hamiltonian. These results show that level spacing statistics is a promising tool to study quasiparticle excitations in flat band superconductors.
\end{abstract}

\maketitle

\section{Introduction}

The discovery of superconductivity in magic-angle twisted bilayer graphene~\cite{caoUnconventionalSuperconductivityMagicangle2018} (MATBG) represents a significant breakthrough for several compelling reasons. It exhibits striking similarities to other unconventional superconductors, such as copper- and iron-based materials~\cite{stewartUnconventionalSuperconductivity2017,keimerQuantumMatterHightemperature2015, stewartSuperconductivityIronCompounds2011, scalapinoCommonThreadPairing2012}, with the superconducting phase emerging when doping from correlated insulating phases characterized by various broken symmetries~\cite{hongDetectingSymmetryBreaking2022,astrakhantsevUnderstandingSymmetryBreaking2023, bultinckGroundStateHidden2020}, forming the characteristic domes. Furthermore, MATBG offers distinct advantages due to its simple composition of pure carbon and remarkable tunability~\cite{checkelskyFlatBandsStrange2024, tormaSuperconductivitySuperfluidityQuantum2022, balentsSuperconductivityStrongCorrelations2020}. Unlike conventional doping methods that introduce disorder, doping in MATBG can be achieved via electrostatic gating, mitigating detrimental effects~\cite{heMoirePatterns2D2021, nuckollsMicroscopicPerspectiveMoire2024, adakTunableMoireMaterials2024}. The twist angle between the graphene layers serves as a crucial tuning parameter, allowing the realization of flat bands; it is precisely near the magic angle, where the bandwidth is minimal, that the superconducting phase appears. The exceptionally high ratio between the superconducting critical temperature and the Fermi temperature, one of the highest among known superconductors~\cite{caoUnconventionalSuperconductivityMagicangle2018}, highlights MATBG as a validation of the concept of enhancing critical temperatures by engineering flat bands with nontrivial geometric and topological properties\cite{kopninHightemperatureSurfaceSuperconductivity2011, heikkilaFlatBandsTopological2011, leykamArtificialFlatBand2018, tormaSuperconductivitySuperfluidityQuantum2022, pyykkonenFlatBandTransport2021}. Specifically, a nonzero quantum metric is essential to ensure that the superconducting state in a flat band remains stable against thermal fluctuations~\cite{peottaSuperfluidityTopologicallyNontrivial2015, tormaQuantumMetricEffective2018, tormaSuperconductivitySuperfluidityQuantum2022}. Studying MATBG holds the promise of elucidating the origins of unconventional superconductivity, a mystery that continues to challenge physicists. These insights could pave the way for increasing the superconducting critical temperature of materials, for instance, through the engineering of flat bands.

The general idea of using flat bands for increasing the critical temperature is very promising but inevitably leads to a challenging theoretical many-body problem. A typical approximation involves projecting onto the subspace of a flat band or a set of flat bands, which significantly reduces the Hilbert space dimension~\cite{bravyiSchriefferWolffTransformation2011, huberBoseCondensationFlat2010, tovmasyanEffectiveTheoryEmergent2016,tovmasyanPreformedPairsFlat2018, swaminathanSignaturesManybodyLocalization2023, mollerCorrelatedPhasesBosons2012, pudleinerInteractingBosonsTwodimensional2015, danieliManybodyLocalizationTransition2022, danieliFlatBandFinetuning2024}. This approach is common in the context of the fractional quantum Hall effect~\cite{stormerFractionalQuantumHall1999, murthyHamiltonianTheoriesFractional2003, jainCompositeFermionsHilbert1997, sodemannLandauLevelMixing2013}. However, a large Wannier function overlap means that the resulting projected Hamiltonian includes numerous relevant terms, deviating considerably from the Hubbard model paradigm of strictly local interactions. The quantum metric is important in this context since it roughly measures the minimal possible overlap of the Wannier functions of a band or a set of bands \cite{marzariMaximallyLocalizedGeneralized1997, tovmasyanEffectiveTheoryEmergent2016}. Consequently, projecting onto a flat band with a large quantum metric yields a complex Hamiltonian that is difficult to handle with currently available methods.

A possible way out is to study simple idealized models first and then proceed to tackle more realistic and complicated ones. In two dimensions, one of the simplest lattices with flat bands is the dice lattice~\cite{vidalAharonovBohmCagesTwoDimensional1998, vidalDisorderInteractionsAharonovBohm2001}. Due to the compact nature of the Wannier functions in the dice lattice, the number of terms obtained after the projection on the two lowest flat bands is limited. Moreover, for attractive interactions, the ground state of the projected Hamiltonian is given exactly by the Bardeen-Cooper-Schrieffer~(BCS) wave function~\cite{tovmasyanEffectiveTheoryEmergent2016, swaminathanSignaturesManybodyLocalization2023, herzog-arbeitmanManyBodySuperconductivityTopological2022}, that is,  the wave function describing the superconducting state at the mean-field level.

Contrary to the ground state, very little is known regarding the nature of the excited states in the dice lattice, and in general, in flat band superconductors; in particular, an outstanding question is whether the quasiparticle excitations are localized~\cite{swaminathanSignaturesManybodyLocalization2023, pyykkonenSuppressionNonequilibriumQuasiparticle2023}. Recent results obtained with exact diagonalization have provided evidence that the dice lattice is an instance of a localized superconductor, that is a superconductor in which the quasiparticle excitations are localized, at least at the level of the projected Hamiltonian.  The concept of a localized superconductor was introduced originally in the context of strongly disordered materials~\cite{maLocalizedSuperconductors1985, maStronglyDisorderedSuperfluids1986}, while the results of Ref.~[\onlinecite{swaminathanSignaturesManybodyLocalization2023}] refer to the disorder-free dice lattice. Localization in the latter case is purely a consequence of the band flatness.

As demonstrated in Ref.~[\onlinecite{swaminathanSignaturesManybodyLocalization2023}],  quasiparticle localization in a flat band superconductor manifests as a weak form of ergodicity breaking. While Cooper pairs propagate and thermalize rapidly, quasiparticles with nonzero spin diffuse very slowly and retain memory of the initial state for extended periods, potentially indefinitely. Ergodicity breaking and thermalization in closed quantum systems are topics that have garnered significant attention~\cite{kliczkowskiFadingErgodicity2024, neillErgodicDynamicsThermalization2016, royStrongErgodicityBreaking2020}, particularly since the concept of many-body localization emerged as a theoretical framework~\cite{oganesyanLocalizationInteractingFermions2007, aletManybodyLocalizationIntroduction2018, nandkishoreManyBodyLocalizationThermalization2015, imbrieLocalIntegralsMotion2017, abaninColloquiumManybodyLocalization2019} and was subsequently observed in ultracold gases in optical lattices~\cite{schreiberObservationManybodyLocalization2015, choiExploringManybodyLocalization2016, kondovDisorderInducedLocalizationStrongly2015}. Many-body localization is fundamentally characterized by an extensive number of local integrals of motion (LIOMs) that appear under strong enough disorder~\cite{serbynLocalConservationLaws2013, serbynCriterionManyBodyLocalizationDelocalization2015, oritoNonthermalizedDynamicsFlatband2021, imbrieLocalIntegralsMotion2017, thomsonLocalIntegralsMotion2023, danieliManybodyFlatbandLocalization2020}. Additionally, Hilbert space fragmentation,  a form of weak ergodicity breaking, can occur due to LIOMs, causing the Hilbert space to fracture into many disconnected sectors within which the dynamics remain ergodic~\cite{moudgalyaQuantumManybodyScars2022, chandranQuantumManyBodyScars2023}.

It is highly desirable, yet generally very challenging, to explicitly produce an extensive number of LIOMs for a many-body Hamiltonian~\cite{chandranConstructingLocalIntegrals2015, obrienExplicitConstructionLocal2016}. Achieving this would dramatically reduce the Hilbert space dimension, making it feasible to numerically simulate large systems. As shown in Ref.~[\onlinecite{swaminathanSignaturesManybodyLocalization2023}], for one term in the projected dice lattice Hamiltonian, the LIOMs are explicitly known and take the same form as those in the Creutz ladder and other one-dimensional lattice models with flat bands~\cite{tovmasyanPreformedPairsFlat2018, kunoFlatbandManybodyLocalization2020}. Specifically, the conserved quantities are the parities of the occupation numbers of each Wannier function, meaning that particles can move only if they form pairs with zero total spin, while single unpaired particles with nonzero spin remain immobile, leading to a frozen spin density distribution. However, other terms in the projected dice lattice Hamiltonian do not uphold this strict dynamical constraint, as evidenced by the slowly diffusive spin dynamics. This does not rule out the presence of LIOMs with a more complex structure, which could underlie the observed weak ergodicity breaking in the dice lattice. Investigating the existence of such LIOMs in lattice models with flat bands is a central motivation of this paper.

Given the difficulty of explicitly constructing LIOMs for a generic many-body Hamiltonian, we use a powerful method to establish the integrability of a quantum system: level spacing statistics~\cite{wignerStatisticalDistributionWidths1951, dysonThreefoldWayAlgebraic1962, guhrRandommatrixTheoriesQuantum1998, santosIntegrabilityDisorderedHeisenberg2004, santosLocalizationEffectsSymmetries2010}. Moreover, instead of directly addressing the projected Hamiltonian of the two-dimensional dice lattice, we introduce another model, the \textit{on-site/bond singlet} (OBS) model. This model is designed to faithfully capture the symmetries and dynamics of the dice lattice while being easier to analyze due to its one-dimensional nature and integrability for certain parameter values.

\begin{figure}[ht]
	\centering 
	\includegraphics[scale =1]{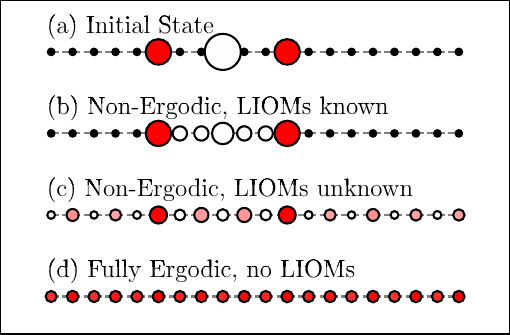}
	\caption{\label{fig:time_evol_chain_20_3_1} Dynamical regimes in the OBS model: (a) Initial state $\ket{\psi(0)} = \ddagup{6}\ddagup{9}\ddagdown{9}\ddagup{12}\ket{\emptyset}$ with an on-site pair on site $l=9$ and two unpaired particles on sites $l=6$ and $12$ ($N_\uparrow, N_\downarrow = 3, 1$) in a chain of length $L=20$ with periodic boundary conditions. The size of the circles denotes the particle density and the color denotes the spin density on a scale from white $S_z=0$ to red $S_z=+1/2$.  (b) Time-averaged asymptotic density obtained from the time evolution with the Hamiltonian $\ham_{\lambda_2,\lambda_3}$ of the OBS model~\eqref{eq:obs_model_Ham} with $\lambda_2=\lambda_3=0$. The on-site pair is trapped in between the localized particles and the spin density is frozen. The LIOMs leading to nonergodic behavior are known  in this case~\cite{tovmasyanPreformedPairsFlat2018}. (c) Same as (b), but for $\lambda_2=0$ and $\lambda_3=1$. The evolution dynamics is partially nonergodic as the asymptotic particle and spin density distributions retain memory of the positions of the unpaired particles. This behavior is similar to that seen in the projected dice lattice Hamiltonian \cite{swaminathanSignaturesManybodyLocalization2023} and indicates the presence of LIOMs, which are not yet known for this model. (d) Same as (b) and (c) but for $\lambda_2=\lambda_3 = 1$. In this case, the asymptotic  particle and spin density distributions are almost uniform, an indication of ergodic behavior.} 
\end{figure}

The Hamiltonian of the OBS model, $\ham_{\lambda_2,\lambda_3}$, given in~\eqref{eq:obs_model_Ham}--\eqref{eq:ham3}, depends on two coupling constants, $\lambda_2$ and $\lambda_3$, which scale terms with direct counterparts in the projected dice lattice Hamiltonian~\cite{swaminathanSignaturesManybodyLocalization2023}. The model exhibits distinct dynamical regimes depending on these parameters, as illustrated in Fig.~\ref{fig:time_evol_chain_20_3_1}. For $\lambda_2=\lambda_3=0$ [Fig.~\ref{fig:time_evol_chain_20_3_1}(b)], the propagating on-site pair remains confined between two localized unpaired particles, which act as impenetrable barriers. This results from the presence of LIOMs enforcing parity conservation of the particle number on each site. Conversely, for $\lambda_2 \neq 0$ and arbitrary $\lambda_3$ [Fig.~\ref{fig:time_evol_chain_20_3_1}(d)], the particle and spin density distributions become nearly uniform, indicating a rapidly thermalizing ergodic regime. The most intriguing regime occurs when $\lambda_2 = 0$ and $\lambda_3 \neq 0$ [Fig.~\ref{fig:time_evol_chain_20_3_1}(c)]. Here, the system exhibits partial nonergodicity, as the particle and spin densities retain memory of the initial state over long times. However, unlike in Fig.~\ref{fig:time_evol_chain_20_3_1}(b), the spin density is not frozen but undergoes slow diffusive dynamics, and the propagating Cooper pair is no longer trapped between the unpaired particles. This occurs because, for $\lambda_3 \neq 0$, the site parities are no longer conserved quantum numbers.

The asymptotic distributions observed in Fig.~\ref{fig:time_evol_chain_20_3_1}(b) closely resemble those found in the projected dice lattice Hamiltonian for an initial state with spin imbalance~\cite{swaminathanSignaturesManybodyLocalization2023}. This similarity is further supported by the eigenvalue spectrum analysis presented below. A key question is whether the partial breaking of ergodicity in the OBS model stems from the presence of LIOMs. One of the main results of this paper suggests that this is indeed the case: the level spacing distribution for $\lambda_2 = 0$ does not change with $\lambda_3$ and remains close to a Poisson distribution, a hallmark of integrable systems. This implies that LIOMs exist for $\lambda_2 = 0$ and arbitrary $\lambda_3 \neq 0$. Furthermore, this raises the possibility of explicitly constructing these LIOMs, as they are likely continuous deformations of those known to exist for $\lambda_2 = \lambda_3 = 0$.

The present paper is structured as follows. The first main result is the identification of a one-dimensional model, the OBS model, which serves as a stepping stone for understanding more complex two-dimensional systems. The OBS model is introduced in Sec.~\ref{sec:obs_model}, while its discrete symmetries and its relation to other models is discussed in Secs.~\ref{sec:symm_ham} and ~\ref{sec:rel_models}, respectively. In Sec.~\ref{sec:excat_diag}, we perform an analysis analogous to that of Ref.~\onlinecite{swaminathanSignaturesManybodyLocalization2023} for the OBS model. Specifically, in Sec.~\ref{sec:spectrum}, we demonstrate that for $\lambda_2=0$, the pattern of degenerate and quasidegenerate states in the OBS model closely resembles that of the projected dice lattice Hamiltonian. The asymptotic density distributions under time evolution are illustrated in Fig.~\ref{fig:time_evol_chain_20_3_1} and further examined in Sec.~\ref{sec:time_evolution}. 

In Sec.~\ref{sec:few_body_problem}, we solve the scattering problem of a two-body bound state impinging on a single unpaired particle for $\lambda_2=0$ and use the results to explain key features of the energy spectrum and out-of-equilibrium dynamics of the OBS model. The level spacing statistics analysis, presented in Sec.~\ref{sec:level_spacing}, provides strong evidence for an extensive number of LIOMs when $\lambda_2=0$ and $\lambda_3 \neq 0$. This effective application of level spacing statistics to the OBS model constitutes the second main result of this paper. Finally, Sec.~\ref{sec:conclusion} provides a summary and gives a perspective on future applications of level spacing statistics in studying quasiparticle localization and weak ergodicity breaking in lattice models with flat bands.

\section{The on-site/bond singlet model}
\label{sec:obs_model}
The OBS model is a one-dimensional fermionic lattice model inspired by the projected dice lattice Hamiltonian, which some of the authors have considered in previous work~\cite{swaminathanSignaturesManybodyLocalization2023}. This model is also an extension of the Hamiltonian obtained by projecting the Hubbard interaction term onto the lowest flat band of the Creutz ladder~\cite{creutzEndStatesLadder1999, takayoshiPhaseDiagramPair2013, tovmasyanGeometryinducedPairCondensation2013, junemannExploringInteractingTopological2017, tovmasyanPreformedPairsFlat2018, kunoFlatbandManybodyLocalization2020}. 
For a concise representation of the Hamiltonian of the 1D OBS model, we introduce sets of operators that are bilinear combinations of fermionic operators, $\hat{d}_{l\sigma}$, $\hat{d}^{\dagger}_{l\sigma}$, where the index $l$ labels different lattice sites on a simple 1D chain, see Fig.~\ref{fig:time_evol_chain_20_3_1}, and $\sigma = \uparrow, \downarrow$ is the spin index. The first set of operators are the on-site spin operators defined as 
\begin{gather}
		\hat{S}_{l}^z = \frac{1}{2}\qty(\hat{d}_{l\uparrow}^\dg\hat{d}_{l\uparrow} - \hat{d}_{l\downarrow}^\dg\hat{d}_{l\downarrow}) \,, \label{eq:Sz} \\
	\hat{S}^+_{l} = \hat{d}_{l\uparrow}^\dg \hat{d}_{l\downarrow} = (\hat{S}^-_{l})^\dg\,,\\
	\hat{S}^x_{l} = \frac{1}{2}\qty\big(\hat{S}^+_{l}+ \hat{S}^-_{l})\,,\qquad \hat{S}^y_{l} = \frac{1}{2i}\qty\big(\hat{S}^+_{l}- \hat{S}^-_{l})\,. \label{eq:SxSy}
\end{gather}
The second set is the on-site pair operators, which describe the creation and annihilation of a pair of particles with opposite spins on the same site \cite{tovmasyanPreformedPairsFlat2018}. These operators obey the same $\mathrm{SU}(2)$ algebra as the on-site spin operators, and are defined as
\begin{gather}
	\hat{B}^z_{l} = \frac{1}{2}(\hat{d}_{l\uparrow}^\dg\hat{d}_{l\uparrow} + \hat{d}_{l\downarrow}^\dg\hat{d}_{l\downarrow} -1)\,, \label{eq:Bz} \\
 	\hat{B}^+_{l} = \hat{d}^\dg_{l\uparrow}\hat{d}^\dg_{l\downarrow} = (\hat{B}^-_{l})^\dg \,, \\
	\hat{B}^x_{l} = \frac{1}{2}\qty\big(\hat{B}^+_{l}+ \hat{B}^-_{l})\,,\qquad \hat{B}^y_{l} = \frac{1}{2i}\qty\big(\hat{B}^+_{l}- \hat{B}^-_{l})\,. \label{eq:BxBy}
\end{gather}
It is important to note that the on-site spin operators and the on-site pair operators commute with each other: $[\hat{S}^\alpha_{l}, \hat{B}_{m}^\beta] = 0$  for $\alpha, \beta = x,y,z$.
The final set of operators are the bond singlet operators that take the form
\begin{equation}
\label{eq:bond_singlet_operator}
\begin{split}
&\hat{B}^+_{\langle l_1,l_2\rangle} = \hat{d}^ \dg_{l_1\uparrow} \hat{d}^\dg_{l_2\downarrow} - \hat{d}^\dg_{l_1\downarrow}\hat{d}^\dg_{l_2\uparrow} \\
&= 	
\hat{d}^ \dg_{l_1\uparrow} \hat{d}^\dg_{l_2\downarrow} + \hat{d}^\dg_{l_2\uparrow}\hat{d}^\dg_{l_1\downarrow} = \qty\big(\hat{B}^-_{\langle l_1,l_2\rangle})^\dg\,. 
\end{split}
\end{equation}
This operator creates a pair of particles with zero total spin (a singlet) delocalized over two lattice sites (the symbol $\langle l_1, l_2\rangle$ denotes an unordered pair of lattice sites). Indeed, it can be observed from (\ref{eq:bond_singlet_operator}) that the state $\hat{B}^+_{\langle l_1,l_2\rangle} \ket{\emptyset}$ is antisymmetric under the exchange of the spin. The $\hat{B}_{\langle l_1, l_2\rangle}^+$ are called \textit{bond} singlets operators since $\langle l_1, l_2\rangle$ will always denote a pair of nearest-neighbor sites ($l_1 = l_2\pm 1$) in the following, therefore it is useful to regard the delocalized singlets as living on the bonds of the simple 1D chain.

The Hamiltonian of the 1D OBS model can be written as
\begin{equation}
\label{eq:obs_model_Ham}
\ham_{\lambda_2, \lambda_3} = \ham_1 + \lambda_2 \ham_2 + \lambda_3 \ham_3\,,
\end{equation}
where each term $\ham_{i = 1,2,3}$ has a direct counterpart in the dice lattice projected Hamiltonian~\cite{swaminathanSignaturesManybodyLocalization2023}.  
Using the sets of operators introduced above, the three terms in the Hamiltonian take the following forms:
\begin{align}
    \ham_1&=
    -A\sum_{j}\hat B^+_j\hat B^-_j 
    -4\sum_{j}\Bigg[\qty(\hat B^z_j 
    +\frac{1}{2})\qty(\hat B^z_{j+1}+\frac{1}{2}) \nonumber \\ &\qquad \qquad +  \frac{1}{2}\qty(\hat{B}^+_{j}\hat{B}^-_{j+1}+\hat{B}^+_{j+1}\hat{B}^-_{j})\Bigg] \nonumber 
    \\ &\quad
    +4\sum_{j=0}^{N-1}\qty[ \hat S^z_j \hat S^z_{j+1}+\frac{1}{2}\qty(\hat S^+_j \hat{S}^-_{j+1}+\hat{S}^-_{j} \hat{S}^+_{j+1})], \label{eq:ham1}
    \\ 
    \ham_2 &=-\sum_{j} \qty[ \hat{B}^+_{\langle j, j+1\rangle}\hat{B}^-_{\langle j+1, j+2\rangle} + H.c.], \label{eq:ham2}
    \\
    \ham_3 &=-\sum_{j} \qty[ (\hat{B}^+_{\langle j, j+1\rangle} - \hat{B}^+_{\langle j-1, j\rangle})\hat{B}^-_j + H.c.]. \label{eq:ham3}
\end{align}
The parameters $\lambda_i$ in~\eqref{eq:obs_model_Ham} are introduced to study the effect of the individual terms $\ham_i$ on the dynamics (see Fig.~\ref{fig:time_evol_chain_20_3_1}) and integrability of the system.

The Hamiltonian $\ham_1$ describes the hopping of on-site pairs (singlets) on the 1D chain and the spin exchange interaction between particles localized on neighboring sites. The first term in $\ham_1$, proportional to $A$, is a 
chemical potential term since $\sum_j \hat{B}_j^+\hat{B}_j^-$ is the number operator of on-site singlets. The last term describes the antiferromagnetic Heisenberg exchange interaction of the on-site spins \cite{tovmasyanEffectiveTheoryEmergent2016}, which can be written in the form $\hat{\vb{S}}_j\cdot\hat{\vb{S}}_{j+1}$ with $\hat{\vb{S}}_j = (\hat{S}_j^x, \hat{S}_j^y, \hat{S}_j^z)^T$. The remaining terms  $\qty\big(\hat{B}_j^z+\frac{1}{2})\qty\big(\hat{B}_{j+1}^z+\frac{1}{2})$  and $\frac{1}{2}\qty\big(\hat{B}^+_{j}\hat{B}^-_{j+1}+\hat{B}^+_{j+1}\hat{B}^-_{j})$ in~\eqref{eq:ham1} represent the nearest-neighbor interaction  and the hopping  of the on-site singlets on the 1D chain, respectively. They can be combined and expressed as an isotropic Heisenberg exchange term $\hat{\vb{B}}_j\cdot\hat{\vb{B}}_{j+1}$ for the pseudospin operators $\hat{\vb{B}}_j=(\hat{B}_j^x, \hat{B}_j^y, \hat{B}_j^z)^T$. 
The second term $\ham_2$ in the 1D OBS model Hamiltonian~\eqref{eq:ham2} describes the hopping of bond singlets along the 1D chain. The last term $\ham_3$ encodes the process through which on-site pairs are converted into bond singlets and vice versa.

\subsection{Symmetries}
\label{sec:symm_ham}

To understand the 1D OBS model, it is useful to identify and resolve the various discrete symmetries that exist in the model.
Therefore, the discrete symmetries of the 1D OBS model are listed in the following. It is important to note that the presence of some of these symmetries is dependent on the parameters $\lambda_2$ and $\lambda_3$. Often the notation in which $\hat{X}$ denotes an operator and $X$ its associated eigenvalue or quantum number is used.

\begin{itemize}
    \item[\textit{(a)}]\textit{Translational symmetry:} The 1D OBS model is translationally invariant, namely the Hamiltonian commutes with the translation operator $\hat{T}$, whose action on the fermionic operators 
    $\hat{d}_{l\sigma}$ is
    \begin{equation}\label{eq:trans_sym}
        \hat{T} \hat{d}_{l,\sigma} \hat{T}^{\dagger} = \hat{d}_{l+1,\sigma}. 
    \end{equation}
    The wave vector $k$ determines the eigenvalues of the operator $\hat{T}$, which take the form $e^{ik}$ since $\hat{T}$ is a unitary operator. The possible values of the wave vector for a chain with $L$ sites are $k = \frac{2\pi l}{L}$ with $l$ an integer in the interval $[0, L-1]$. 
    \item[\textit{(b)}]\textit{Reflection (parity) symmetry:} When $\lambda_3 = 0$, the 1D OBS model possesses reflection symmetry described by the operator $\hat{R}$,
    \begin{equation}
        \hat{R} \hat{d}_{l,\sigma} \hat{R}^{\dagger} = \hat{d}_{L-l+1,\sigma}.
    \end{equation}
    The reflection center can be chosen at any site or bond center due to the presence of the translational symmetry mentioned above. The reflection operator has the property $\hat{R}^2 = 1$, therefore the only possible eigenvalues are $R = \pm 1$. Since performing the reflection operation reverses the direction of the wave vector, it is possible to find simultaneous eigenstates of $\hat{T}$ and $\hat{R}$ only for the  wave vectors $k = 0$  or $k=\pi$ that are invariant under reflections. The inclusion of $\ham_3$, the conversion term between bond singlets and on-site pairs, breaks the reflection symmetry otherwise present in the system. 
    More specifically, the symmetry is broken by the opposite signs of the amplitudes associated to the processes of bond singlet creation in the forward $(\hat{B}^+_{\langle j, j+1\rangle}\hat{B}^-_j)$ and backward $(- \hat{B}^+_{\langle j-1, j\rangle}\hat{B}^-_j)$ directions in~\eqref{eq:ham3}. It would have been possible to choose the same amplitude for both the forward and backward directions and thus preserve reflection symmetry, however, doing so would lead to the breaking of other symmetries that are important for our purposes, in particular, the conservation of the total pseudospin discussed below.
    
    \item[\textit{(c)}]\textit{Spin-rotation symmetry:} The 1D OBS model exhibits $\mathrm{SU}(2)$ spin-rotation symmetry. This indicates that the Hamiltonian commutes with the components of the total spin operator, i.e.,
    \begin{equation}\label{eq:Salpha}
        [\ham_i, \hat{S}^\alpha] = 0,
    \end{equation}
     for $i = 1, 2, 3$ and $\alpha = x, y, z$. The components of the total spin operator are defined as $\hat{S}^\alpha = \sum_l \hat{S}^\alpha_l $ with the single-site spin operators $\hat{S}^\alpha_l$ previously defined in~\eqref{eq:Sz} and \eqref{eq:SxSy}. The total spin operator is given by $\hat{S}^2 = (\hat{S}^x)^2 + (\hat{S}^y)^2 + (\hat{S}^z)^2$. 
     Thus, the total spin $S$, defined in terms of the eigenvalue $S(S+1)$ of the total spin  operator $\hat{S}^2$, and the total spin along the $z$-axis $S^z$ are good quantum numbers. 
     
    \item[\textit{(d)}]\textit{Pseudospin symmetry:} Unlike the components of the spin operator $\hat{S}^\alpha$, not all components of the total pseudospin operators $\hat{B}^{\alpha} = \sum_l \hat{B}^{\alpha}_l $ defined using Eqs.~\eqref{eq:Bz} and \eqref{eq:BxBy} commute with the complete Hamiltonian of the 1D OBS model. Only the $\hat{B}^z$ component is conserved for arbitrary values of $\lambda_{i=2,3}$, namely,
        \begin{equation}
        [\ham_i, \hat{B}^z] = 0
    \end{equation}
    for $i = 1, 2, 3$. Note that $\hat{B}^z =(\hat{N}-L)/2$ is equivalent to the total particle number operator $\hat{N}=\sum_{j\sigma}\hat{d}_{j\sigma}^\dagger\hat{d}_{j\sigma}$ up to a constant. Together with the conservation of $\hat{S}^z$ from Eq. \eqref{eq:Salpha}, we see that the number of particles for each spin component is conserved, with $N_\uparrow = B^z + S^z+L/2$ and $N_\downarrow = B^z - S^z+L/2$.

    While $\ham_1$ and $\ham_3$ possess full pseudospin symmetry, the bond singlet hopping term $\ham_2$ breaks it since
    \begin{align}
        [\ham_1 \, , \hat{B}^+] &= -(A+4)\hat{B}^+, \label{eq:SGA_H1}     
        \\
        [\ham_2, \hat{B}^+] &\not \propto \hat{B}^+, \label{eq:comm_H2_Bp} 
        \\
        [\ham_3, \hat{B}^+] &= 0.
        \label{eq:SGA_H3}  
    \end{align}
    As a consequence of the first and last commutation relations, one has
    \begin{equation}
    [\ham_1, \hat{B}^2] = [\ham_3, \hat{B}^2]=0\,.
    \end{equation}
    On the other hand, $\ham_2$ does not commute with $\hat{B}^2$.
    The commutation relation between $\ham_1$ and $\hat{B}^+$ is called a dynamical symmetry or spectrum generating algebra  (SGA), which takes the generic form $[\ham \, , \hat{X}] = \xi \hat{X}$ with $\xi \neq 0$~\cite{barutDynamicalGroupsMass1965, bucaNonstationaryCoherentQuantum2019, moudgalyaQuantumManybodyScars2022}.
    A SGA can be used to generate towers of equally spaced energy eigenstates. Indeed, if $\ket{\psi_0}$ is an eigenstate of $\ham$ with eigenvalue $\varepsilon_0$, then $\hat{X}^n\ket{\psi_0}$ are also eigenstates with eigenvalues $\varepsilon_0 + n\xi$ for $n$ a positive integer, provided that $\hat{X}^n\ket{\psi_0}\neq 0$~\cite{moudgalyaQuantumManybodyScars2022}.
    
    \item[\textit{(e)}]\textit{Local integrals of motion:} The presence of local integrals of motion is established for $\ham_1$ as it is identical to that of the previously studied Creutz ladder \cite{tovmasyanPreformedPairsFlat2018, kunoFlatbandManybodyLocalization2020} if $A=4$. The LIOMs are given by the total spin operators at each site $\hat{S}^2_l$, thus the total spin quantum number at site $l$ takes value $S_l = 1/2$ if one particle is present and $S_l=0$ otherwise. As discussed extensively in Ref.~\cite{tovmasyanPreformedPairsFlat2018}, this implies that particles can move within the lattice only if they form an on-site pair. Unpaired particles remain localized to their sites as shown in Fig.~\ref{fig:time_evol_chain_20_3_1}(b). However, these LIOMs do not commute with the complete Hamiltonian due to the presence of the terms $\ham_2$ and $\ham_3$.
    For the following, it is important to note that the LIOMs operators $\hat{S}^2_l$ along with the translation operator $\hat{T}$ generate a non-Abelian symmetry algebra due to the obvious relation $\hat{T} \hat{S}^2_l \hat{T}^{\dagger} = \hat{S}^2_{l+1}$.

\end{itemize}

In Sec.~\ref{sec:level_spacing}, we analyze the level spacing statistics of $\ham_{\lambda_2,\lambda_3}$, particularly how it evolves when varying $\lambda_2$. Therefore, the symmetry blocks of the Hamiltonian of the 1D OBS model are labeled by the quantum numbers $(N_{\uparrow}, N_{\downarrow}, S, k, [R])$. The square brackets indicate that the eigenvalue $R$ of the reflection operator is included only when applicable. When performing exact diagonalization, we always consider symmetry blocks with  wave vector $k=0$. Only when $\lambda_3 = 0$, is it possible to specify the reflection symmetry eigenvalue, which is set to $R=+1$. To avoid accounting for the spin inversion symmetry $S^z\to -S^z$ (not listed above), we only consider the spin imbalanced case $N_\uparrow - N_\downarrow = 2S^z\neq 0$.

\subsection{Related models}
\label{sec:rel_models}

The Hamiltonian of the 1D OBS model is analogous to that of the projected dice lattice Hamiltonian, albeit in one dimension. To faithfully capture the physics of the latter one, it is essential that the OBS model maintains as many of its symmetries as possible. In this section, we compare and contrast the OBS model with the projected dice lattice Hamiltonian, as well as the closely related Creutz ladder in particular with respect to the symmetry properties.

As mentioned above, the term $\ham_1$~\eqref{eq:ham1} is identical (for $A=4$) to the Hubbard interaction term projected onto one flat band of the Creutz ladder  \cite{tovmasyanPreformedPairsFlat2018}.
The corresponding term in the Hamiltonian obtained by projecting the Hubbard interaction term onto the two lowest flat bands of the dice lattice is denoted by $\ham_{\rm tri.}$ in Ref.~\cite{swaminathanSignaturesManybodyLocalization2023} and is responsible for the spin dynamics and the hopping of the on-site singlets within the triangular lattice formed by the Wannier functions centers. 
The symmetry properties of $\ham_1$ and $\ham_{\rm tri}$ are essentially the same, in particular, they both possess an extensive number of LIOMs of the form $\hat{S}^2_{l}$, where $l$ labels different Wannier functions in the case of the dice lattice.
It should be noted that, in contrast to the projected Hamiltonian of the dice lattice, the 1D OBS model is not derived as the projection of a Hubbard term onto the flat band of any lattice model, at least to our knowledge.

\begin{figure}[ht]
    \centering
    \includegraphics[width=\linewidth]{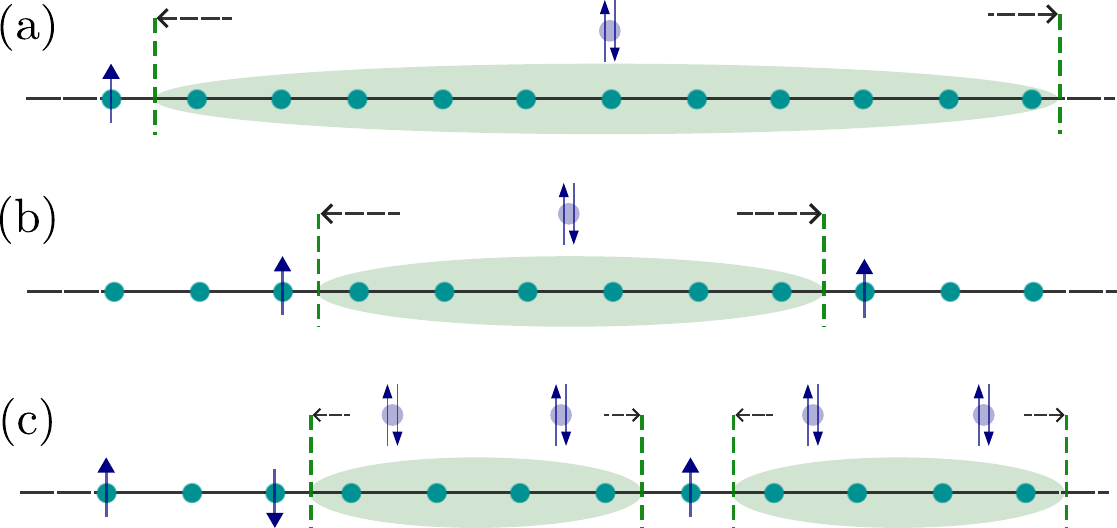}
    \caption{\label{fig:spin_LIOM} Illustration of possible eigenstates of the integrable Hamiltonian $\ham_1 = \ham_{0,0}$~\eqref{eq:ham1} for a chain of length $L=12$ under periodic boundary conditions containing different numbers of particles. (a) A possible configuration for $(N_\uparrow, N_\downarrow = 2, 1)$ with one section where the delocalized pair is present. (b) A possible state for $(N_\uparrow, N_\downarrow = 3, 1)$. (c) A possible state for  $(N_\uparrow, N_\downarrow = 6, 5)$ with two equal-sized sections in which the on-site pairs can propagate. In each section the Hamiltonian reduces to an isotropic Heisenberg model for the pseudospin degree of freedom with open boundary conditions and additional boundary fields, see~\eqref{eq:H1_three-body}.}
    
\end{figure}

An important advantage of working in one dimension is that $\ham_1$ is integrable~\cite{tovmasyanPreformedPairsFlat2018}. This is a key reason for introducing the 1D OBS model. The integrability of the Hamiltonian $\ham_1$ can be demonstrated as it possesses an extensive number of LIOMs, ${\hat{S}^2_l}$, as described in Sec.~\ref{sec:symm_ham}. When the eigenvalues of the LIOMs are specified, the positions of the unpaired particles are fixed at the sites $l$ on the chain where $S_l=1/2$. These localized particles partition the chain into sections where the Hamiltonian simplifies to independent isotropic Heisenberg models for either the spin or pseudospin degrees of freedom. Figure ~\ref{fig:spin_LIOM} illustrates some examples of these partitions for a chain of length $L=12$ with different numbers of particles. In general, these Heisenberg models are subject to open boundary conditions with additional boundary fields. However, when no unpaired particles are present or when unpaired particles fill the whole chain, the model reduces to a single pseudospin or spin Heisenberg model with periodic boundary conditions. The isotropic Heisenberg model under both open and periodic boundary conditions is known to be integrable \cite{caoDiagonalBetheAnsatz2013}. As $\ham_1$ comprises of several independent integrable models, it is itself integrable.

The term $\ham_2$ is analogous to $\ham_{\rm kag.}$ in the projected dice lattice Hamiltonian, while $\ham_3$ is  comparable to $\ham_{\rm tri.-kag.}$; see Ref.~[\onlinecite{swaminathanSignaturesManybodyLocalization2023}]. The projected dice lattice Hamiltonian possesses both spin and pseudospin rotation symmetries. In particular, it is shown in Ref.~[\onlinecite{swaminathanSignaturesManybodyLocalization2023}] that $[\ham_{\rm kag.}+\ham_{\rm tri.-kag.}, \hat{b}^\dg]  \propto\hat{b}^\dg$, where the operator $\hat{b}^\dg$ in the same reference is the analog of $\hat{B}^+$ in the case of the dice lattice. Note that the SGA does not hold separately for $\ham_{\rm kag.}$ and $\ham_{\rm tri.-kag.}$ in the projected dice lattice Hamiltonian. By choosing the appropriate relative sign of the forward and backward terms in $\ham_{3}$, it is possible to preserve the SGA in~\eqref{eq:SGA_H1} at the cost of breaking reflection symmetry, as discussed in Sec.~\ref{sec:symm_ham}.
In contrast to $\ham_{\rm tri.-kag.}$, we were unable to adapt the bond singlet hopping term $\ham_{\rm kag.}$ to one dimension, while preserving the conservation of total pseudospin. Hence, the commutator $[\ham_2, \hat{B}^+]$ is not compatible with the SGA satisfied by the other terms in OBS model~\eqref{eq:comm_H2_Bp}. Nevertheless, we find it useful to include $\ham_2$ in the 1D OBS model as an integrability breaking perturbation (see Sec.~\ref{sec:level_spacing}).

 Due to the terms $\ham_{\rm kag.}$ and $\ham_{\rm tri.-kag.}$, which describe the motion of bond singlets and the conversion between on-site  and bond singlets within the projected dice lattice, it is unclear whether an extensive number of LIOMs exists in this model. Indirect evidence for the presence of LIOMs has been found in Ref.~[\onlinecite{swaminathanSignaturesManybodyLocalization2023}] by analyzing the time-evolution dynamics. Similarly, we cannot make an assertion regarding the existence of LIOMs in the OBS model due to the presence of the terms associated with bond singlets, $\ham_2$ and $\ham_3$. The aforementioned terms, unlike $\ham_1$, do not possess an obvious set of LIOMs. 
 
 In fact, the main goal of the present paper is to explore whether such conserved quantities exist for the 1D OBS model. The results shown in the following sections, particularly the level spacing analysis presented in Sec.~\ref{sec:level_spacing}, suggest that $\ham_{0,\lambda_3}$ is integrable for any value of $\lambda_3$. The Hamiltonian presumably possesses an extensive number of LIOMs that are comparable but not identical to those of the Creutz ladder Hamiltonian $\ham_1$. In contrast, integrability breaks for $\ham_{\lambda_2\neq 0, \lambda_3}$ and no LIOMs are present.

\section{Exact diagonalization results}
\label{sec:excat_diag}

In this section, we analyze the energy spectrum and the time-evolution dynamics of the 1D OBS model. We provide numerical results for the energy spectrum of the Hamiltonian when the  interaction terms $\ham_2$ and $\ham_3$ are present or absent.  Additionally, the time evolution dynamics for the same Hamiltonians are shown for a representative initial state that includes both on-site pairs and unpaired particles. 
This analysis is essential to establish the 1D OBS model as a platform to understand the physics of its two-dimensional counterpart, the dice lattice, and, more generally, the class of Hamiltonians obtained by projecting an Hubbard interaction term onto a flat band.

All of the numerical results in the following sections are
obtained using the exact diagonalization package \textsc{QuSpin} \cite{weinbergQuSpinPythonPackage2017, weinbergQuSpinPythonPackage2019}. The free parameter in $\ham_1$~\eqref{eq:ham1} is fixed to $A=10$. For the calculations in this section, periodic boundary conditions are used, which amounts to the identification $\hat{d}_{L+1,\sigma} \equiv \hat{d}_{1,\sigma}$ in~\eqref{eq:ham1}--\eqref{eq:ham3}. 
With the \textsc{QuSpin} package, it is possible to directly construct the Hamiltonian on the subspace specified by the set of good quantum numbers $(N_{\uparrow}, N_{\downarrow}, k, [R])$,  discussed in Sec.~\ref{sec:symm_ham}. However, with \textsc{QuSpin} it is not possible to specify the values of the total spin $S$ and pseudospin $B$. To resolve these symmetries,  their corresponding operators are added to the Hamiltonian multiplied by sufficiently large coefficients, i.e., $\ham_{\lambda_2,\lambda_3} + \alpha \hat{S}^2 + \beta \hat{B}^2$~\cite{poilblancPoissonVsGOE1993}.

\begin{figure}[ht]
    \centering 
    \includegraphics[scale=1]{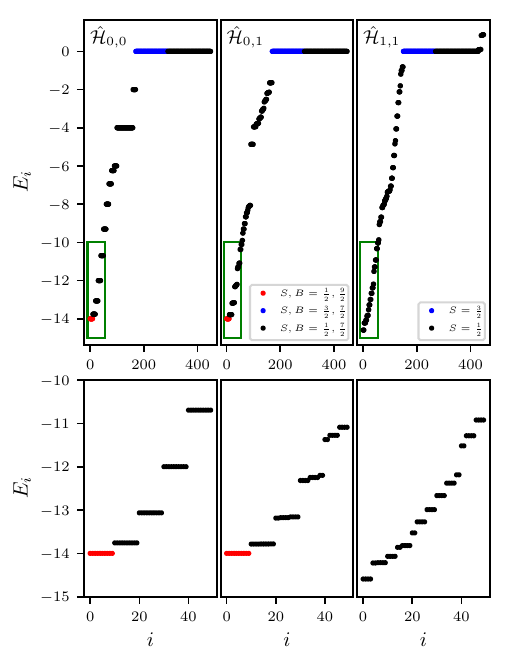}
    \caption{\label{fig:ev_2_1} Energy spectrum of the Hamiltonian $\ham_{\lambda_2,\lambda_3}$ of the 1D OBS model obtained from exact diagonalization for a chain of length $L = 10$, particle numbers $N_\uparrow, N_\downarrow = 2,1$, and $A=10$, see~\eqref{eq:ham1}. In the three columns, the energy eigenvalues for different choices of the  coupling constants $\lambda_2,\,\lambda_3$ are shown. The lower portion of each  energy spectrum indicated by the rectangle in the upper row is magnified in the lower row. The quantum numbers for total spin $S$ and total pseudospin $B$ are indicated by different colors according to the legend. In the case $\lambda_2\neq 0$, only the total spin is a good quantum number.} 
\end{figure}

\subsection{Energy Spectrum}
\label{sec:spectrum}

In Fig. \ref{fig:ev_2_1}, the eigenvalues for the Hamiltonian $\ham_{\lambda_2, \lambda_3}$ with three particles $(N_{\uparrow}, N_{\downarrow} =2, 1)$ are shown for a chain of length $L=10$. The first column depicts the eigenvalues of the Hamiltonian $\ham_{0,0}$,  where the lowest $L$ eigenstates are exactly degenerate. The excited states in this case can also be grouped into sets of the same energy with the size of the sets being exactly equal to the ground state degeneracy. A similar feature remains upon the addition of the term $\ham_3$, as seen in the middle column of Fig.~\ref{fig:ev_2_1}. As one can observe from the bottom row of  Fig.~\ref{fig:ev_2_1}, where the smallest eigenvalues are highlighted, the lowest excited states of $\ham_{0,1}$ can be grouped in sets of approximately the same energy, with the size of the sets still equal to their ground state degeneracy. This structure in the excited state spectrum can be understood from the existence of LIOMs for $\ham_1$ (Sec.~\ref{sec:symm_ham}). The eigenvalues $S_l(S_{l}+1)$ of the operators $\hat{S}^2_l$, i.e., the total spin on the $l$-th lattice site, are good quantum numbers. The lowest lying states have $S_l = 1/2$ for a single site, with $S_l = 0$ elsewhere since there exists only one unpaired particle [see Fig.~\ref{fig:spin_LIOM}(a)]. Given a simultaneous eigenstate of $\ham_1$ and all the spin operators $\hat{S}^2_l$, one can obtain new distinct eigenstates that have the same energy by repeatedly applying the translation operator. These new eigenstates are distinct since the localized unpaired particle is moved to a different site as a consequence of the nontrivial commutation relation $\hat{T} \hat{S}^2_l \hat{T}^{\dagger} = \hat{S}^2_{l+1}$. Thus, the degeneracy of the lowest eigenstates is at least equal to the the length of the chain $L$. The observation of a similar structure in the excited state spectrum of $\ham_{0,1}$ and $\ham_{0,0}$ is an indication that the LIOMs survive for $\lambda_3 \neq 0$. This hypothesis is further corroborated in Sec.~\ref{sec:level_spacing}, where it is shown 
that also the level spacing distributions of $\ham_{0,0}$ and $\ham_{0,1}$ are remarkably similar. Indeed, the level spacing statistics are an unbiased method for characterizing the similarity of the energy spectra of two Hamiltonians~\cite{guhrRandommatrixTheoriesQuantum1998, giraudProbingSymmetriesQuantum2022}. 
 
The aforementioned structure is lost in the energy spectrum of $\ham_{1,1}$ as a consequence of the addition of the bond singlet hopping term $\ham_2$ to the system. As shown in the rightmost column of Fig.~\ref{fig:ev_2_1}, there is no clear organization of the eigenstates in quasidegenerate sets, which we interpret as evidence that  $\ham_{1,1}$ does not possess local conserved quantities. Again, the level spacing statistics analysis of Sec.~\ref{sec:level_spacing} further supports this scenario. In the dice lattice, for comparison, quasidegenerate sets of states similar to those in the middle column of Fig.~\ref{fig:ev_2_1} are found when both $\ham_{\rm kag.}$ and $\ham_{\rm tri.-kag.}$ are present at the same time, while the cases in which only one of the two terms is included has not been considered in Ref.~[\onlinecite{swaminathanSignaturesManybodyLocalization2023}]. As a reminder, $\ham_{\rm kag.}$ and $\ham_{\rm tri.-kag.}$ correspond to $\ham_2$ and $\ham_3$, respectively.
Note that the ground state degeneracy of $\ham_{1,1}$ is significantly reduced compared to the cases where $\lambda_2 =0$. The exact ground state degeneracy is due to the SGA~\eqref{eq:SGA_H1}--\eqref{eq:SGA_H3}, which is not satisfied when $\lambda_2\neq 0$~\eqref{eq:comm_H2_Bp}.
Indeed, from the SGA and the fact that the single-particle state $\hat{d}^\dg_{j\sigma}\ket{\emptyset}$ is a zero energy eigenstate, it follows that the three-particle states
\begin{equation}
\label{eq:exact_three_body}
\ket{j,\sigma} = \hat{B}^+\hat{d}_{j\sigma}^\dagger\ket{\emptyset}
\end{equation}
are also eigenstates~\cite{swaminathanSignaturesManybodyLocalization2023} and in fact form the degenerate ground state manifold according to the numerical results.

\begin{figure}[ht]
    \centering 
    \includegraphics[scale = 1]{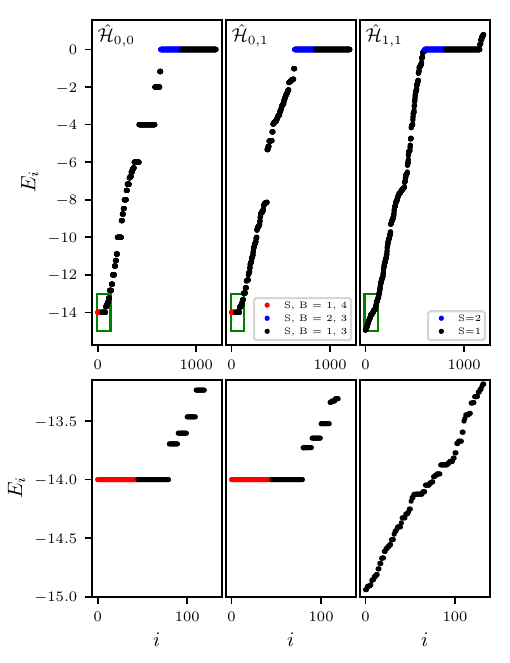}
    \caption{\label{fig:ev_3_1} Same as in Fig.~\ref{fig:ev_2_1}, for the case $N_\uparrow, N_\downarrow = 3, 1$. When $\lambda_2 = 0$, the ground state degeneracy observed  is larger than $\binom{L}{2}$ as expected for two unpaired particles on a chain.  Equal-sized sets of degenerate ($\lambda_3 = 0$) or quasi-degenerate excitated states ($\lambda_3 \neq 0$)  are present in the spectrum when $\lambda_2 = 0$, which is interpreted as a signature of the presence of LIOMs.} 
\end{figure}

The energy spectra in the case of $N_\uparrow,\,N_\downarrow =3,\,1$ for the same values of $\lambda_2$ and $\lambda_3$ are shown in Fig.~\ref{fig:ev_3_1}. Again, the lowest excited states of $\ham_{0,1}$ are organized into quasidegenerate sets, obtained by slightly lifting the degeneracy of the exactly degenerate sets of excited states with multiplicity $L$ as in the case of $\ham_{0,0}$. The presence of these quasidegenerate excited states for $\ham_{0,1}$ indicates the survival of the LIOMs. On the other hand, the total degeneracy of the ground state manifold is larger than $\binom{L}{N_{\uparrow} - N_{\downarrow}}$, the number of possible ways of placing $N_{\uparrow}-N_{\downarrow}$ unpaired particle in the chain. This formula for the ground state degeneracy holds in the case of the two-dimensional dice lattice~\cite{swaminathanSignaturesManybodyLocalization2023} but fails for the OBS model. The reason for the larger ground state degeneracy in one dimension can be understood from Fig.~\ref{fig:spin_LIOM}. The unpaired particles partition the chain into sections where the Hamiltonian reduces to the isotropic Heisenberg model, as discussed in Sec.~\ref{sec:symm_ham}. By solving explicitly the three-body scattering problem (Sec.~\ref{sec:three_body_problem}), it is shown that the ground states manifold of $\ham_{0,\lambda_3}$ is spanned by states of the form
\begin{align}\label{eq:gs_3_1_a}
\ket{j_1,\sigma_1;j_2,\sigma_2}_1 &= \hat{B}_{[j_1+1, j_2-1]}^+\hat{d}^\dagger_{j_1 \sigma_1} \hat{d}^\dagger_{j_2 \sigma_2}\ket{\emptyset}, \\
\label{eq:gs_3_1_b}
\ket{j_1,\sigma_1;j_2,\sigma_2}_2 &= \hat{B}_{[j_2+1, L+j_1-1]}^+\hat{d}^\dagger_{j_1 \sigma_1} \hat{d}^\dagger_{j_2 \sigma_2}\ket{\emptyset},
\end{align}
where $\hat{B}_{[l_1, l_2]}^+$ is defined as
\begin{align}
    \hat{B}_{[l_1,l_2]}^+ &= \sum_{l =l_1 }^{l_2 }\hat{B}_{l}^+\,.
\end{align}
Here, the convention of labeling lattice sites by integers modulo  $L$ is used, thus the on-site pair creation operators are identified according  to $\hat{B}_l^+\equiv \hat{B}_{L+l}^+$. The operator $\hat{B}_{[l_1,l_2]}^+$ creates a particle in a plane wave state with $k=0$ restricted to the lattice sites $l=l_1,\dots,l_2$.
An example of the states in~\eqref{eq:gs_3_1_a} and \eqref{eq:gs_3_1_b}  is illustrated in Fig.~\ref{fig:spin_LIOM}(b). It follows that the ground state degeneracy for $N_\uparrow, N_\downarrow = 3, 1$ on a chain subject to periodic boundary conditions is $L + 2\qty\big[\binom{L}{2} - L]$.

\subsection{Evolution Dynamics}
\label{sec:time_evolution}

\begin{figure}[ht]
    \centering 
    \includegraphics[scale=1]{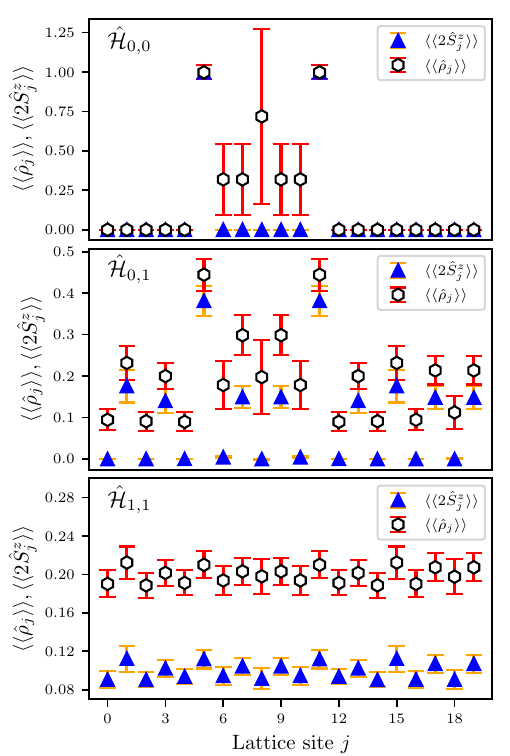}        \caption{\label{fig:time_evol_20_3_1} 
    Time-averaged particle and spin densities obtained from the time evolution with the OBS model Hamiltonian $\ham_{\lambda_2, \lambda_3}$. Time evolution is performed for a chain of length $L=20$ with four particles $(N_\uparrow, N_\downarrow = 3, 1)$ initially in the state~\eqref{eq:initial_state}. Different variations of the Hamiltonian are shown in each row, from top to bottom: $\ham_{0,0}, \ham_{0,1}, \ham_{1,1}$. The time average for the particle and spin densities at each site are calculated from $t_0 = 500$ to $t_1 = 2000$ \eqref{eq:time-average_observable}. The error bars indicate the standard deviation \eqref{eq:stddev} of each quantity over the same time interval.} 
\end{figure}

In this section, we consider the particle and spin densities obtained  from the time evolution with the OBS model Hamiltonian $\ham_{\lambda_2,\lambda_3}$ of a four-particle initial state of the form 
\begin{equation}
\label{eq:initial_state}
    \ket{\psi(0)} = \ddagup{6}\ddagup{9}\ddagdown{9}\ddagup{12} \ket{\emptyset}\,.
\end{equation}
The initial state consists of an on-site pair and two unpaired particles, as illustrated in Fig.~\ref{fig:time_evol_chain_20_3_1}(a). The time-averaged particle and spin densities for $\ham_{0,0}$, $\ham_{0,1}$, and $\ham_{1,1}$ are shown schematically in rows (b), (c), and (d), respectively, and also in Fig.~\ref{fig:time_evol_20_3_1}, where in addition the standard deviation is also reported. Here, the time average of an observable is defined as 
\begin{equation} \label{eq:time-average_observable}
    \langle \langle \hat{O} \rangle \rangle = \frac{1}{t_1 -t_0} \int_{t_0}^{t_1} \langle \hat{O}(t) \rangle dt \, ,
\end{equation}
where $\langle \hat{O}(t) \rangle$ is the expectation value of an observable $\hat{O}$ at time $t$. The initial cutoff time $t_0$ is large  enough to ensure that the expectation value $\ev*{\hat{O}}$ is close to its long-time asymptotic value and $t_1$ is the final time. 
The standard deviation shown in Fig.~\ref{fig:time_evol_20_3_1} is computed as
\begin{equation} \label{eq:stddev}
    \sigma_{\hat{O}} = \sqrt{\langle {\langle \hat{O} \rangle}^2 \rangle - {\langle \langle \hat{O} \rangle \rangle}^2}.
\end{equation}
The time averages of the spin and particle densities are calculated from $t_0=500$ to $t_1=2000$. 

The results for the different Hamiltonians showcase various degrees of coexistence of ergodic and nonergodic behavior in the system. In the case of $\ham_{0,0}$, the on-site pair exhibits ergodic behavior whereas the unpaired particles are strictly localized on their respective sites. This behavior is expected due to the presence of LIOMs for $\ham_1$, previously discussed in Sec.~\ref{sec:symm_ham}

In the case of the Hamiltonian $\ham_{0,1}$, both the on-site pair and unpaired particles propagate along the whole chain, but the signature of the initial positions of the unpaired particles are observed through the increased spin and particle densities at their respective sites. This behavior was previously observed and studied for the 2D case in the dice lattice \cite{swaminathanSignaturesManybodyLocalization2023}. The unique scattering process responsible for the observed quasiparticle diffusion will be analyzed further in Sec.~\ref{sec:few_body_problem}. In particular, it will become apparent why the spin density is concentrated on the even sites of the chain.

However, with the addition of the bond singlet hopping term $\ham_2$, both the on-site pair and unpaired particles appear to exhibit fully ergodic behavior since the information about the initial state is completely lost. This is the behavior expected in chaotic systems. The chaotic nature of the Hamiltonian $\ham_{1,1}$ is confirmed by the analysis of the level spacing statistics in Sec.~\ref{sec:level_spacing}.

\section{The few-body problem in the on-site/bond singlet model}\label{sec:few_body_problem}

In this section, we investigate analytically the few-body problem for the OBS Hamiltonian $\ham = \ham_1 + \lambda_3 \ham_3$. We do not consider the term $\ham_2$ in this section since it does not commute with the pseudospin operator $\Bplus{}$~\eqref{eq:comm_H2_Bp},  moreover including it would not provide significant insights into the physics of the OBS model.

\subsection{Two-body problem}\label{sec:two_body_problem}

\begin{figure}[h]
    \includegraphics[scale = 1]{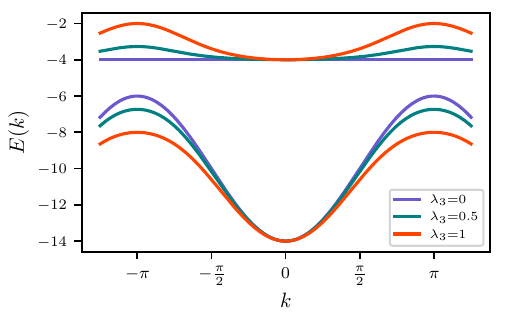}
    \caption{Dispersion of the OBS model $\ham = \ham_1 + \lambda_3 \ham_3$ for different values of $\lambda_3$ with lattice constant $a=1$. For $\lambda_3 = 0$ the bond singlet states~\eqref{eq:fourier_basis_2} form a flat band, which acquires a finite bandwidth for increasing $\lambda_3$. The dispersion around $k=0$ is only slightly affected by the parameter $\lambda_3$. }
    \label{fig:two-body_dispersion}
\end{figure}

We first investigate the two-body problem for the Hamiltonian $\ham = \ham_1 + \lambda_3 \ham_3$. A basis of the nontrivial two-body subspace for this Hamiltonian is the set
of states containing a single one-site or bond singlet:
\begin{align}
    \qty{\Bplus{j}\vac,\, \Bijplus{j}{j+1}\vac \,\bigg|\, j = 0,\dots,L-1}\,. 
\end{align}
To take advantage of translational invariance, it is convenient to form plane wave linear combinations
\begin{align}
    &\ket{\phi_1(k)} = \frac{1}{\sqrt{L}}\sumj e^{i kja}\Bplus{j}\vac, \label{eq:fourier_basis_1}
    \\
    & \ket{\phi_2(k)} = \frac{1}{\sqrt{2L}}\sumj e^{i kja}\Bijplus{j}{j+1}\vac, \label{eq:fourier_basis_2}
\end{align}
where $a$ is the lattice constant for the chain. 
These states form an orthonormal basis of the nontrivial subspace, thus the two-body problem is solved by diagonalizing 
\begin{align} \label{eq:H_eff_two_body}
    H_{\rm 2b}(k) =  
    \begin{bmatrix}
-\qty(A+4\cos ka) & -\lambda_3 \sqrt{2}\qty(1-e^{-ika}) \\
-\lambda_3 \sqrt{2}\qty(1-e^{ika}) & -4 
\end{bmatrix}.
\end{align}
The eigenstates of $H_{\rm 2b}(k)$ are denoted as $\ket{\Phi_l(k)}$, where $\ket{\Phi_1(k)}$ becomes the on-site pair state $\ket{\phi_1(k)}$ in the $\lambda_3 \to 0$ limit,  while $\ket{\Phi_2(k)}$ reduces to the bond singlet plane wave $\ket{\phi_2(k)}$. 
The same occurs for any $\lambda_3$ when $k=0$ since the off-diagonal elements in $H_{\rm 2b}(k=0)$ vanish.
The dispersions $E_l(k)$ of the two-body bound states are shown in Fig.~\ref{fig:two-body_dispersion}.
With increasing  $\lambda_3$, the flat band of bond singlets hybridizes with the on-site singlets and acquires a nonzero bandwidth. At the same time the bandwidth of the lower band is decreased. On the other hand, around $k=0$, the band structure is essentially unchanged due to the fact that the off-diagonal terms in $H_{\rm 2b}(k)$ tend to zero for $k \to 0$. 

\subsection{Three-body problem}\label{sec:three_body_problem}

Here, we consider the three-body problem for the OBS model. An orthonormal basis of the three-body subspace for $N_\uparrow,\,N_\downarrow = 2,1 $ is given by
\begin{align}\label{eq:basis_set_three_body}
   \ket{l,m,n} = \hat{d}_{l\up}^\dg \hat{d}_{m\dw}^\dg \hat{d}_{n\up}^\dg\ket{\emptyset} \qq{with} l > n\,. 
\end{align}
Many of these states are already eigenstates of the Hamiltonian, for instance in the case in which no two particles are adjacent to each other or reside on the same site. Thus, it is possible to select a subset of relevant three-body states on which the Hamiltonian acts nontrivially. The specific choice depends on whether the interaction term $\ham_3$ is present or not.
First, we study the case in which the on-site/bond singlet conversion term is absent and the Hamiltonian reduces to $\ham_1$. In this case, the bond singlets do not propagate, thus only states that contain one on-site pair and one unpaired particle need to be considered.  In the notation of~\eqref{eq:basis_set_three_body}, these states have the form
\begin{align} \label{eq:restricted_basis_set}
&\ket{n,n,m} \qq{if} m < n,  \nonumber \\ &\ket{m,n,n} \qq{if}  m > n\,.
\end{align}
Without loss of generality, we set $m = 0$ and introduce the following more compact notation
\begin{equation}\label{eq:restricted_basis_set2}
	\ket{n} = \hat{B}_n^+\hat{d}^\dg_{0,\up}\ket{\emptyset} = \begin{cases}
		\ket{n,n,0} \qq{if} n > 0\,\\
		-\ket{0,n,n} \qq{if} n < 0\,.
	\end{cases}
\end{equation}
The state $\ket{0} = \hat{B}_0^+\hat{d}^\dg_{0,\up}\ket{\emptyset} = 0$ does not exist as a consequence of the Pauli exclusion principle. 

The three-body Hamiltonian $H_{\rm 3b}$ is obtained by projecting the many-body Hamiltonian $\ham_1$ on the subspace spanned by the above states and reads
\begin{equation} \label{eq:H1_three-body}
\begin{split}
			H_{\rm 3b} =  &- A\sum_{n\neq0}\dyad{n} - 2\qty\big(\dyad{1} + \dyad{-1})  \\ &- 2\sum_{n \neq 0,-1}\qty\big(\dyad{n+1}{n} + \dyad{n}{n+1})\,. 
\end{split}
\end{equation}
Thus, one simply needs to solve the scattering problem of a particle against a potential consisting of an infinite potential barrier at site $l=0$ and a finite potential well at sites $l=\pm 1$, given by the second term in $H_{\rm 3b}$.
No transmission of the on-site pair is possible through the potential barrier, therefore the only unknown in the scattering problem is  the phase of the reflected wave.  
This can be readily found by solving the Schr\"odinger equation and one obtains the following scattering states separately for the right- and left-hand sides of the chain
\begin{gather}
    \label{eq:right_scattering_wf}
    \begin{split}
    \ket{\psi_+(k)} &= \sum_{n > 0}\qty\big(e^{ikn}+e^{ik(1-n)})\ket{n} \\
    &= 2e^{ik/2}\sum_{n > 0}\cos\qty[k\qty(n-\frac{1}{2})]\ket{n}\,,
    \end{split}
    \\
    \label{eq:left_scattering_wf}
    \begin{split}
    \ket{\psi_-(k)} &= \sum_{n < 0}\qty\big(e^{ikn}+e^{-ik(1+n)})\ket{n} 
    \\
    &=2e^{-ik/2}\sum_{n < 0} \cos\qty[k\qty(n+\frac{1}{2})]\ket{n}\,,
    \end{split}
\end{gather}
with energy $E(k) = -A - 4\cos ka$. Note that for  $k = \pi$ there is no solution since $\ket{\psi_\pm(\pi)} = 0$. 

The three-body Hamiltonian for a finite chain of length $L$ with periodic boundary conditions is obtained from~\eqref{eq:H1_three-body} by identifying the states $\ket{n} \equiv \ket{L+n}$. The same Hamiltonian describes the situation in which two unpaired particles are located at sites $n=0$ and $n = L$  of a chain of size $L' > L$. In both cases, shown, respectively, in Figs.~\ref{fig:spin_LIOM}(a) and~\ref{fig:spin_LIOM}(b),  the on-site pair  can visit only the sites with $0 < n < L$.
The eigenstates in a finite chain are found by imposing that the wave function takes simultaneously the two forms~\eqref{eq:right_scattering_wf} and~\eqref{eq:left_scattering_wf} up to an arbitrary phase factor. This is equivalent to imposing suitable boundary conditions on the left- and right-hand sides, respectively.
In this way, one finds the following stationary states: 
\begin{equation}\label{eq:three_body_solution}
	\ket{\psi_k} = \sum_{n = 1}^{L-1} \cos\qty [k\qty(n-\frac{1}{2})]\ket{n}\,,\quad k = \frac{\pi m}{L-1}\,,
\end{equation}
with $m = 0, 1, .... L-2$.
These form a complete basis of states for a single on-site pair confined between two unpaired particles. A very peculiar feature of this solution is that the plane wave with $k=0$ is an eigenstate and its energy is independent of $L$. This is a consequence of the finely tuned boundary potential in $H_{\rm 3b}$, that is, the term $-2(\dyad{1}+\dyad{L-1})$ in~\eqref{eq:H1_three-body}. For any other choice of the magnitude of the boundary potential the energy would depend on the chain length and the density distribution would not be uniform.
Therefore, the result that states of the form~\eqref{eq:exact_three_body} are exact eigenstates of the many-body Hamiltonian has been recovered from the explicit solution of the three-body problem.

In fact, the explicit solution gives additional information that cannot be obtained from the SGA alone, namely, the fact that the states in~\eqref{eq:three_body_solution} are also eigenstates of the four-body problem in which an on-site pair is confined between two unpaired particles [see Fig.~\ref{fig:spin_LIOM}(b)]. In the special case $k=0$, these are the states in~\eqref{eq:gs_3_1_a} and  \eqref{eq:gs_3_1_b} that compose the ground state manifold in the case $N_\uparrow,\,N_\downarrow = 3,1$. 

When $\lambda_3 \neq 0$ the many-body Hamiltonian $\ham = \ham_1 +\lambda_3 \ham_3$ is projected on the subspace spanned by the states in~\eqref{eq:basis_set_three_body}. 
The resulting three-body Hamiltonian is denoted as $H_{\rm 3b} = H_1 + \lambda_3H_3$, where the first term $H_1$ reads 
 \begin{widetext}
\begin{gather} \label{eq:H1_three-body_2}
\begin{split}
H_1 &=  -A\sum_{\substack{m,n\\ m < n}}\dyad{n,n,m}	- A\sum_{\substack{m,n\\ m > n}}\dyad{m,n,n} \\
&-2\sum_{\substack{m,n\\m <n}}\qty\big(\dyad{n+1,n+1,m}{n,n,m} + \dyad{n,n,m}{n+1,n+1,m}) \\
&-2\sum_{\substack{m,n\\m >n}}\qty\big(\dyad{m,n,n}{m,n-1,n-1} + \dyad{m,n-1,n-1}{m,n,n})\\
&-2\sum_{n}\qty\big(\dyad{n+1,n+1,n} + \dyad{n+1,n,n}) \\
&-2\sum_{\substack{m,n\\m <n}}\qty\big(\ket{n+1,n,m}+\ket{n,n+1,m})
\qty\big(\bra{n+1,n,m}+\bra{n,n+1,m})\\
&-2\sum_{\substack{m,n\\m >n}}\qty\big(\ket{m,n,n-1}+\ket{m,n-1,n})
\qty\big(\bra{m,n,n-1}+\bra{m,n-1,n})\,,
\end{split}
\end{gather}
while for the on-site/bond singlet conversion term $H_3$, one obtains the following representation
\begin{gather}
\label{eq:H_3_projected_three-body}
\begin{split}
H_3 = &-\sum_{\substack{n,m\\ m < n}}\qty\big[\qty\big(\ket{n,n,m}-\ket{n+1,n+1,m})\qty\big(\bra{n+1,n,m}+\bra{n,n+1,m})+\mathrm{H.c.}]\\
&-\sum_{\substack{n,m\\ m > n}}\qty\big[\qty\big(\ket{m,n-1,n-1}-\ket{m,n,n})\qty\big(\bra{m,n,n-1}+\bra{m,n-1,n})+\mathrm{H.c.}]\,. \\
\end{split}
\end{gather}

Here, we are particularly interested in solving the scattering problem in which one two-body bound state propagates (from left to right) and interacts with a a single unpaired particle. The initial site for the unpaired particle is set at $m = 1$. Whereas for $\lambda_3 = 0$ it is sufficient to restrict the basis to states where the unpaired particle remains fixed at the initial site, see \eqref{eq:restricted_basis_set2}, this is not possible  when $\lambda_3 \neq 0$.  Consequently, we consider a larger subspace spanned by states in the union  $S_1\cup S_2$ of two sets:
\begin{equation} \label{eq:inv_set1}
S_1 = \big\lbrace\ket{1,n,n}, \, \ket{1,n,n-1}, \, \ket{1,n-1,n} \! \! \! \! \qq{for} \! \! \! \! n < 1\big\rbrace\,, 
\end{equation}
\begin{equation} \label{eq:inv_set2}
S_2 = \big\lbrace\ket{n,n,-1}, \, \ket{n+1,n,-1}, \, \ket{n,n+1,-1} \! \! \! \! \qq{for} \! \! \! \! n > -1\big\rbrace\,.
\end{equation}
\end{widetext}
These two sets are not disjoint, with the state $\ket{1,0,-1}$ being an element of both. The subspace spanned by $S_1\cup S_2$ is an invariant subspace for $H_{\rm 3b}$, as one can verify directly from~\eqref{eq:H_3_projected_three-body}. It is interesting to note that states of the form $\ket{n,n,0}$ and $\ket{0,-n,-n}$ are not elements of the invariant subspace. This implies that the interaction of the propagating two-body bound state with the unpaired particle can displace the latter by two sites, but not by only one site, and  explains why the spin density is nonzero only on the even sites  in Fig.~\ref{fig:time_evol_chain_20_3_1}(c) (see also Fig.~\ref{fig:time_evol_20_3_1}). 
The main effect of the on-site/bond singlet conversion term is to create the possibility of tunneling through an unpaired particle by means of the following process:
\begin{equation}
\label{eq:tunnel_proc}
	\ket{1,0,0} \to \ket{1,0,-1} \to \ket{0,0,-1}\,.	
\end{equation}
Note that this is a second order process since it requires two on-site/bond singlet conversions.

To construct a scattering solution of the three-body problem, the following Ansatz is considered:
\begin{equation}
\label{eq:3b_ansatz}
    \ket{\psi_{l,k}} = \sum_{\substack{\ket{n_1,n_2,n_3}\in S_1 \cup S_2 }}\psi_{l,k}(n_1,n_2,n_3)\ket{n_1,n_2,n_3},
\end{equation}
where\begin{gather}
    \label{eq:3b_wf_1}
    \psi_{l,k}(1,n,n) = \braket{n,n}{\Phi_l(k)} + r_l(k)\braket{n,n}{\Phi_l(-k)} ,  \\
    \label{eq:3b_wf_2}
    \begin{split}
    \psi_{l,k}(1,n,n-1) &= \braket{n,n-1}{\Phi_l(k)} \\
    &+ r_l(k)\braket{n,n-1}{\Phi_l(-k)}, 
    \end{split}
    \\
    \label{eq:3b_wf_3}
    \psi_{l,k}(1,n-1,n) = \psi_{l,k}(1,n,n-1),
\end{gather}
with $n <0$, and
\begin{gather}
    \label{eq:3b_wf_4}
    \psi_{l,k}(n,n,-1) = t_l(k)\braket{n,n}{\Phi_l(k)},  \\
    \label{eq:3b_wf_5}
    \psi_{l,k}(n+1,n,-1) = t_l(k)\braket{n+1,n}{\Phi_l(k)},  \\
    \label{eq:3b_wf_6}
    \psi_{l,k}(n,n+1,-1) = \psi_{l,k}(n+1,n,-1)
\end{gather}
for $n >0$.
The wave function~\eqref{eq:3b_ansatz} is a linear combination of  incident $\ket{\Phi_l(k)}$ and reflected $r_l(k)\ket{\Phi_l(-k)}$ waves on the left-hand side and a single transmitted wave $t_l(k)\ket{\Phi_l(k)}$ on the right-hand side. The reflection and transmission amplitudes are denoted with $r_l(k)$ and $t_l(k)$, respectively, while $\braket{m,n}{\Phi_l(k)}$ are the expansion coefficients of the two-body bound  state wave functions of Sec.~\ref{sec:two_body_problem} in the two-body basis $\ket{m,n} = \hat{d}^\dg_{m\uparrow}\hat{d}_{n\downarrow}^\dg\ket{\emptyset}$. 
The conditions in~\eqref{eq:3b_wf_3}
and~\eqref{eq:3b_wf_6} are a consequence of the fact that the wave function of a bond singlet is spatially symmetric. Note how, in the above wave function, the unpaired particle is located at site $l=1$ when the two-body bound state is on the left-hand side. On the other hand, the unpaired particle is found at $l=-1$ if the bound state is on the right-hand side in agreement with~\eqref{eq:tunnel_proc}. 

\begin{figure}[h]\includegraphics[scale=1]{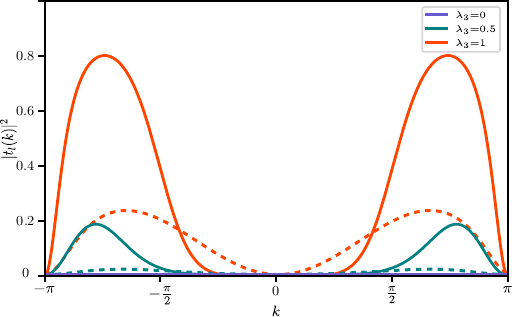} \caption{\label{fig:three_body_transmission} Transmission coefficient $\abs{t_l(k)}^2$ from the three-body problem in the OBS model. $\abs{t_1(k)}^2$ is represented by the solid line and $\abs{t_2(k)}^2$ by the dashed line for different values of $\lambda_3$. When $\lambda_3 = 0$, $t_l(k)$ is completely suppressed. Note that the transmission coefficient is suppressed around $k=0$.}
\end{figure}

The only unknowns in~\eqref{eq:3b_ansatz} are the coefficients that appear on the left-hand sides of~\eqref{eq:3b_wf_1}--\eqref{eq:3b_wf_6} for $n=0$ together with the reflection and transmission amplitudes. By requiring that the Schr\"odinger equation $H_{\rm 3b}\ket{\psi_{l,k}} = E_l(k)\ket{\psi_{l,k}}$ is satisfied, one can determine the wave function completely. In particular the transmission coefficient $\abs{t_l(k)}^2 = 1-\abs{r_l(k)}^2$ is shown in Fig.~\ref{fig:three_body_transmission}. Exactly at $k=0$ the transmission coefficient vanishes since on-site singlets are decoupled from bond singlets as discussed in Sec.~\ref{sec:two_body_problem}, thus the solution for $\lambda_3 = 0$ is recovered. It also means that the four-body states in~\eqref{eq:gs_3_1_a} and \eqref{eq:gs_3_1_b} are exact degenerate ground states also when the term $\ham_3$ is included in the Hamiltonian.

Beside explaining the larger than expected degeneracy of the ground state discussed in Sec.~\ref{sec:spectrum}, the main merit of the analytical solution of the three-body scattering problem is to reveal how an unpaired particle can move around the chain by repeatedly colliding with a two-body bound state. The basic process is represented in~\eqref{eq:tunnel_proc} and leads to the diffusion of the spin density observed in Figs.~\ref{fig:time_evol_chain_20_3_1} and~\ref{fig:time_evol_20_3_1} for $\lambda_3 \neq 0$ and $\lambda_2 = 0$. 

\section{Level spacing statistics of the on-site/bond singlet model}
\label{sec:level_spacing}

\begin{figure*}[ht]
    \centering 
    \includegraphics[scale = 1]{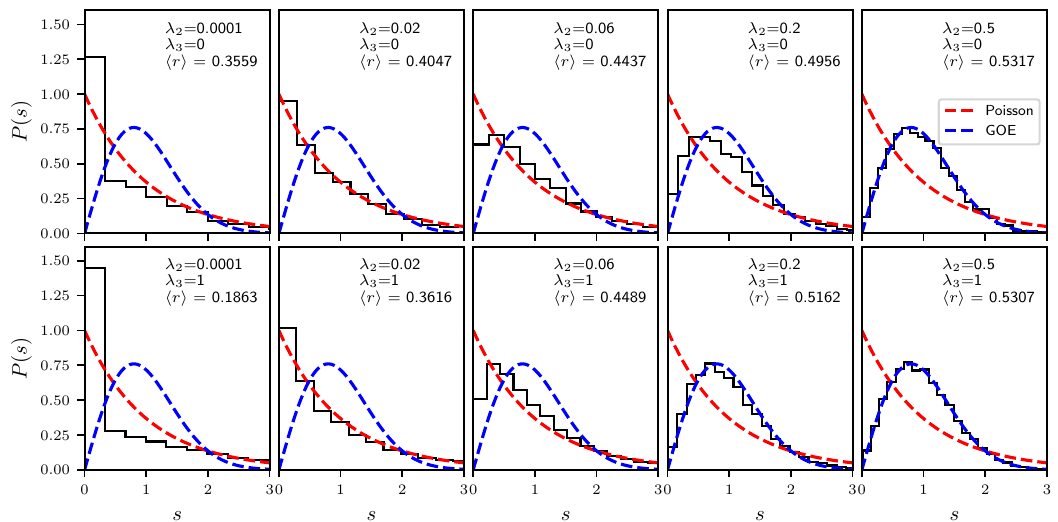}
    \caption{\label{fig:lss_pbc_lambda_3} Level spacing distribution for the 1D OBS model subject to periodic boundary conditions. Level spacings were calculated for a chain of length $L=12$ for various values of $\lambda_2$, while $\lambda_3 = 0$ in the top row and $\lambda_3= 1$ in the bottom row. The Poisson and GOE distributions are indicated by the red and blue dashed lines, respectively. In the top row, the Hamiltonian is diagonalized in the subspace with quantum numbers $(N_\uparrow, N_\downarrow, S, k, R) = (5, 6, 1/2, 0, 1)$ (subspace dimension $\mathcal{N} = 6294$). The same choice of quantum numbers is employed in the bottom row with the only difference that the quantum number $R$ cannot be specified since the reflection operator does not commute with $\ham_3$. Then the subspace dimension is $2\mathcal{N}$. Note that the distributions in the two rows are similar for identical values of $\lambda_2$, irrespective of the value of $\lambda_3$.
    } 
\end{figure*}


A common approach to understand the statistical behavior of quantum many-body systems is the random matrix theory \cite{mehtaRandomMatrices2004} originally developed by Wigner \cite{wignerStatisticalDistributionWidths1951}. He proposed that the gaps within the energy spectrum of heavy nuclei should have the same probability distribution of the spacings between the eigenvalues of a random matrix. If this so-called Wigner surmise holds true, universal statistical properties of the energy spectrum  of a system with strong interactions and complex behavior can be obtained simply by identifying its symmetries~\cite{ guhrRandommatrixTheoriesQuantum1998}. There are three random matrix symmetry classes with different level spacing distributions: the complex Hermitian, real symmetric, or the quaternion self-dual symmetry classes \cite{dysonThreefoldWayAlgebraic1962,zirnbauerSymmetryClassesRandom2006}. The OBS model is time-reversal symmetric, thus the relevant symmetry class is the one of real symmetric random matrices, also called the Gaussian orthogonal ensemble (GOE) in random matrix theory.

\begin{figure*}[ht]
    \centering 
    \includegraphics[scale = 1]{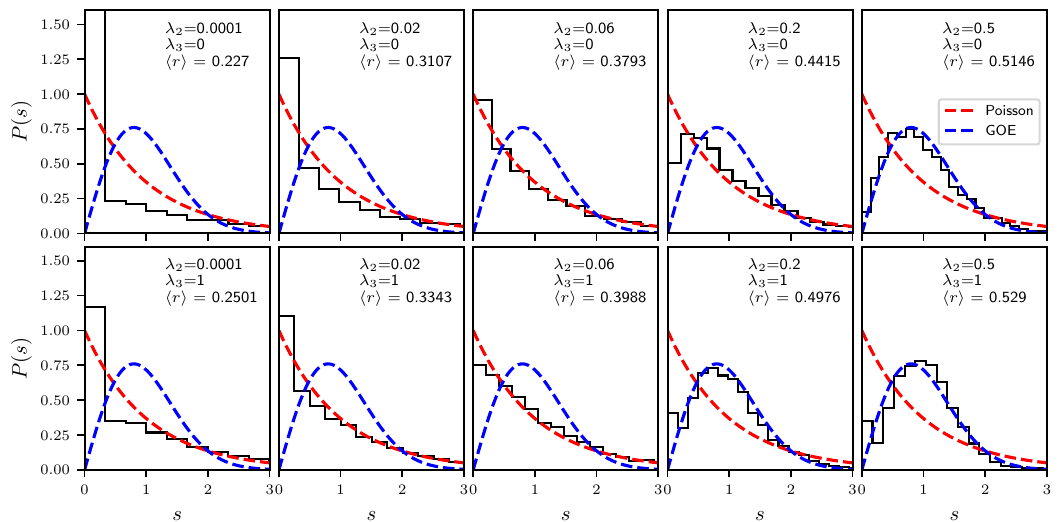}
    \caption{\label{fig:lss_obc_lambda_3}
    Level spacing distribution in the 1D OBS model for a chain of length $L=10$ with open boundary conditions. The values of the parameters $\lambda_2$, $\lambda_3$ are as in Fig.~\ref{fig:lss_pbc_lambda_3}. In the top row, the subspace quantum numbers are $(N_\uparrow, N_\downarrow, S, k, R) = (4, 5, 1/2, 0, 1)$ and the subspace dimension is $\mathcal{N} = 3656$. For the bottom row, $(N_\uparrow, N_\downarrow, S, k) = (4, 5, 1/2, 0)$ and the subspace dimension is $2\mathcal{N}$. Note that the distributions in the two rows are similar for identical values of $\lambda_2$, irrespective of the value of $\lambda_3$.
    } 
\end{figure*}

To compare the statistical properties of the energy spectrum of a Hamiltonian with the universal results from random matrix theory, we consider their unfolded eigenvalues. Unfolded eigenvalues are the renormalized eigenvalues obtained from rescaling the energy spectrum by the local average level spacing or local density of states. These unfolded eigenvalues are dimensionless and their local density of states is equal to one throughout the spectrum \cite{guhrRandommatrixTheoriesQuantum1998, brodyRandommatrixPhysicsSpectrum1981, bruusEnergyLevelStatistics1997}. In general, the unfolding procedure is performed by finding the system specific mean level density or by parametrizing a numerically obtained level density in the terms of smooth functions such as polynomials. Here, we utilize an unfolding procedure similar to that of Ref.~[\onlinecite{santosIntegrabilityDisorderedHeisenberg2004}], in which we first discard $2.5\%$ of the levels from each edge of the spectrum to remove the contributions from the parts of the spectrum where large fluctuations may exist. Next, we compute the level spacings $\delta_n$, i.e., the gaps between adjacent energy levels,
\begin{equation}
\delta_n = E_{n+1} - E_n \geq 0\,,
\end{equation}
where ${E_n}$ are the eigenvalues for the relevant Hamiltonian listed in ascending order. These spacings are split into 30 equal sections, where the size of each bin is dependent on the total number of levels in the portion of the spectrum analyzed. The mean energy level spacing $M$ is then calculated for each section from which we obtain the unfolded level spacings $s_n=\delta_n/M$. The unfolded level spacing distributions for the 1D OBS model with varying contributions of $\ham_2$ and $\ham_3$ are shown in Figs.~\ref{fig:lss_pbc_lambda_3} and~\ref{fig:lss_obc_lambda_3}.

The distribution of these renormalized level spacings for a Hamiltonian gives insight into the integrability of the model. For integrable models possessing an extensive number of conserved quantities, the distribution of the level spacings should follow a Poisson distribution, $P(s) = e^{-s}$. On the contrary, The level spacing distribution of a model that behaves chaotically will follow that of a Gaussian ensemble of random matrices~\cite{guhrRandommatrixTheoriesQuantum1998}. In particular, a chaotic model with time-reversal symmetry, follows the GOE whose unfolded level spacing distribution has the Wigner-Dyson form~\cite{giraudProbingSymmetriesQuantum2022, atasDistributionRatioConsecutive2013},
\begin{equation}
P(s) = \frac{\pi s}{2} \exp(-\frac{\pi s^2}{4})\,.
\end{equation}

Figure~\ref{fig:lss_pbc_lambda_3} presents samples of the level spacing statistics for the 1D OBS model under periodic boundary conditions. The distributions depicted in the top row are for the Hamiltonian $\ham_{\lambda_2, 0}$ for values of $\lambda_2$ in the range $[0.0001, 0.5]$, while the bottom row shows the distribution for the Hamiltonian $\ham_{\lambda_2, 1}$ with the same values of $\lambda_2$. When $ \lambda_2 = \lambda_3 = 0$, one would expect the Poisson distribution due to the integrability of $\ham_{0,0}$. However, one observes that the bin with lowest level spacing $s$ is much larger than expected from the Poisson distribution when $\lambda_2 \lesssim 0.02$. 

The anomaly in the level spacing distribution of the 1D OBS model for $\lambda_2=\lambda_3 = 0$ can be explained by the presence of the LIOMs, which together with the translation operator generate a non-Abelian symmetry algebra (see Sec.~\ref{sec:symm_ham}). The eigenvalues $S_l$ of the LIOMs determine the positions of the unpaired particles, as explained in Sec.~\ref{sec:rel_models}. Each configuration of unpaired particles divides the system into independent sections in which the Hamiltonian reduces to a simple Heisenberg model for the spin or the pseudospin. Some of these partitions may contain sections with the same length. This is shown in Fig.~\ref{fig:spin_LIOM}(c), illustrating one such configuration on a chain of $L=12$ with two equal-sized sections of length four in which on-site pairs propagate. The energy eigenvalues of two identical sections with the same number of particles are the same, leading to a large number of degenerate states that manifest as the unusually large bin with lowest $s$ in the histogram of the unfolded eigenvalues. However, this degeneracy is lifted in the presence of $\ham_2$, a consequence of the fact that the LIOMs are broken by this term. Indeed, with increasing $\lambda_2 \gtrsim 0.2$, the level spacing distribution approaches that of the GOE, which is the level spacing distribution of a real symmetric random matrix. 

The key result illustrated in the lower row of Fig.~\ref{fig:lss_pbc_lambda_3} is that the large number of degenerate states persists even with the addition of the on-site/bond singlet conversion term $\ham_3$, i.e., when $\lambda_3 = 1$. The addition of this interaction term does not qualitatively change the statistical properties of the level spacings for any value of $\lambda_2$. We interpret this as evidence that LIOMs exist even in the presence of $\ham_3$. However, the operators $\hat{S}_l^2$ do not commute with $\ham_{0,1}$ and, consequently, the LIOMs for $\ham_{0,\lambda_3 \neq 0}$ necessarily have a different form, for which there is no explicit expression at present. 

Similar behavior in the level spacing statistics is observed for the OBS model subject to open boundary conditions as shown in Fig.~\ref{fig:lss_obc_lambda_3}. Again, for small values of $\lambda_2$, the number of level spacings with lowest $s$ is much larger than in the usual Poisson distribution. This observation can also be explained by the existence of many partitions of the chain with equal-sized sections when LIOMs exist. Moreover, similar to the OBS model under periodic boundary conditions, there is an evident transition to chaotic behavior for $\lambda_2 \gtrsim 0.2$. Crucially, the presence of $\ham_3$ does not appear to alter the statistics significantly, furthering the evidence that the LIOMs are preserved by this term. 

An alternative method to analyze the statistical properties of the energy spectrum of a quantum many-body system was proposed by Oganesyan and Huse \cite{oganesyanLocalizationInteractingFermions2007} in terms of gap ratio $r_n$ of three consecutive energy levels, defined as 
\begin{equation}
      \label{eq:gapratio}
      r_n =\frac{\mathrm{min}(\delta_n,\delta_{n-1})}{\mathrm{max}(\delta_n,\delta_{n-1})}.
\end{equation}
Since the ratio of consecutive levels is independent of the local density of states, utilizing the gap ratio instead of the level spacings eliminates the prerequisite knowledge of the density of states. Consequently, this removes the need to perform an unfolding procedure.
The distribution of the gap ratio $P(r)$ has been used in recent work, 
since it allows for a more precise comparison with experiments as compared to the level spacing distribution $P(s)$ \cite{oganesyanLocalizationInteractingFermions2007, santosLocalizationEffectsSymmetries2010, santosOnsetQuantumChaos2010, palManybodyLocalizationPhase2010, rigolQuantumChaosThermalization2010}. 

The quantity $r_n$ is always, by nature of its construction, a positive number in the interval $[0,1]$. The mean value of the gap ratio in the case of an integrable system with level spacing statistics described by the Poisson distribution is $\langle r \rangle \approx 0.3863$. The numerical estimate for the mean gap ratio of the GOE, which describes the statistics of a chaotic system, is $\langle r \rangle \approx 0.5359$ \cite{atasDistributionRatioConsecutive2013,giraudProbingSymmetriesQuantum2022}. In the following, the mean gap ratio of the OBS model for different values of $\lambda_2$ and $\lambda_3$ is compared against these two limits. 

\begin{figure}[ht]
    \centering 
    \includegraphics[scale = 1]{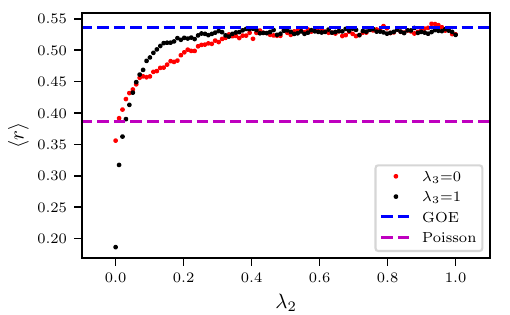}
    \caption{\label{fig:r2_H2_pbc} Mean gap ratio $\langle r \rangle$ of the OBS model Hamiltonian $\ham_{\lambda_2, \lambda_3}$ with periodic boundary conditions. The parameters are as in Fig.~\ref{fig:lss_pbc_lambda_3}. The blue and purple dashed lines indicate the mean gap ratios for the GOE and Poisson distribution, respectively. The  value of $\langle r \rangle$ near $\lambda_2 = 0$ is below the expected value for the Poisson distribution. Note that the inclusion of $\ham_3$ does not affect the general behavior of the mean gap ratio.} 
\end{figure}

\begin{figure}[ht]
    \centering 
    \includegraphics[scale = 1]{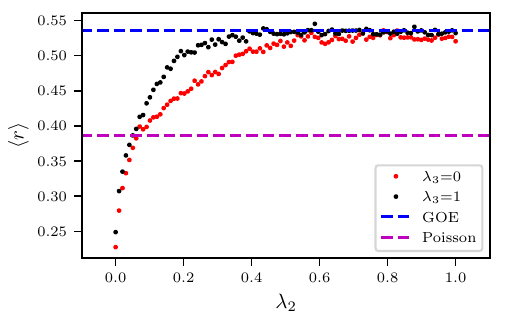}
    \caption{\label{fig:r2_H2_obc}Same as Fig. \ref{fig:r2_H2_pbc} but for a chain  subject to open boundary conditions. 
    The parameters are the same as in Fig.~\ref{fig:lss_obc_lambda_3}.
    Similar to Fig. \ref{fig:r2_H2_pbc}, the addition of the term $\ham_3$ to the Hamiltonian does not change the qualitative behavior of the mean gap ratio of the OBS model as a function of $\lambda_2$.
    }
\end{figure}

In Fig. \ref{fig:r2_H2_pbc}, the mean gap ratio of the OBS model  close to half filling is presented as a function of $\lambda_2$. The mean gap ratio is calculated for periodic boundary conditions in the two cases where the contribution of the on-site/bond singlet conversion term $\ham_3$ is switched off ($\lambda_3 = 0$) or switched on ($\lambda_3 = 1$). Regardless of the value of $\lambda_2$, the mean gap ratio transitions from a value lower than the Poisson distribution, which characterizes integrable models, to that of the GOE which describes chaotic behavior. The mean gap ratio quickly approaches the expected value for the GOE for values of $\lambda_2 \gtrsim 0.2$. An interesting feature when $\lambda_2 \sim 0$ is that the mean gap ratio is smaller than that of the Poisson distribution. This feature appears as a consequence of the large number of  degenerate or quasi-degenerate energy levels as shown in Fig.~\ref{fig:lss_pbc_lambda_3}. This, in turn, is a manifestation of the special kind of LIOMs commuting with $\ham_1$, as explained above. It is possible that the size of the chain studied could impact the number of degenerate states, with larger chains leading to a growing number of degeneracies. However, this finite-size effect can not be confirmed due to computational limits. Remarkably, this behavior is independent of the presence of $\ham_3$, as visible in Fig.~\ref{fig:r2_H2_pbc}. There is only a minor qualitative difference as the system approaches chaotic behavior relatively quickly when $\lambda_3 = 1$ as compared to when $\lambda_3 = 0$.    

In the case of open boundary conditions, the behavior of the mean gap ratio is very similar to the one for periodic boundary conditions. As seen in Fig.~\ref{fig:r2_H2_obc}, the transition of the mean gap ratio from lower than the Poisson distribution to the GOE value is similar to the case with periodic boundary conditions. Similar to the model with periodic boundary conditions, the sub-Poissonian mean gap ratio for the model when $\lambda_2 \sim 0$ furthers the argument for LIOMs for $\ham_3$. Again, the only notable difference in the spectral statistics of the two Hamiltonians $\ham_{\lambda_2,0}$ and $\ham_{\lambda_2,1}$ is that the presence of the term $\ham_3$ accelerates the convergence of the mean gap ratio to the GOE value. This effect can also be observed in Fig.~\ref{fig:lss_obc_lambda_3}.

The anomalous level spacing distribution of the 1D OBS model when $\ham = \ham_{\lambda_2,0}$ for small $\lambda_2$ can be attributed to the local integrals of motion for $\ham_1$ described in Sec.~\ref{sec:symm_ham}. The similar level spacing distribution observed for $\lambda_3=1$ provides evidence for the existence of LIOMs in the presence of the on-site/bond singlet conversion term $\ham_3$. It is clear from the numerical results that the boundary conditions do not affect the LIOMs. However, it is important to stress that the level spacing statistics analysis can only provide indirect evidence for the existence of local conserved quantities. Therefore, it would be preferable to develop a direct method to construct them.

\section{Conclusion and Discussion}
\label{sec:conclusion}

We have introduced the OBS model as a one-dimensional analog of the projected dice lattice Hamiltonian previously studied in Ref.~[\onlinecite{swaminathanSignaturesManybodyLocalization2023}], to better understand the peculiar coexistence of ergodic and nonergodic behavior observed in the latter. The dice lattice with an attractive Hubbard interaction is probably the simplest example of a flat band superconductor in two dimensions. Despite this, very little is known regarding the nature of its excitations; in particular, an outstanding open question is whether the observed partial breaking of ergodicity in the out of equilibrium dynamics is due to the presence LIOMs.

As a first result, it has been shown with exact diagonalization that the energy spectrum and the out-of-equilibrium dynamics of the OBS model are remarkably similar to those of its two-dimensional counterpart when the bond singlet hopping term is not included ($\lambda_2 = 0$). In this way, the OBS model is established as a useful tool to investigate quasiparticle localization in flat band superconductors in the simpler setting of one dimension. For instance, we have been able to solve explicitly the three-body scattering problem, gaining valuable insight on the dynamical processes that govern quasiparticle and spin transport.

A key advancement of this paper is having demonstrated that level spacing statistics is a powerful tool for unveiling the mystery of quasiparticle excitations in flat band superconductors since it confirms and substantially extends the results obtained from the analysis of the energy spectrum and the time dynamics of few particles.
From our point of view, the most important result of the present paper is showing that the level spacing distribution of the Hamiltonian $\ham_{0,\lambda_3}$ is essentially independent of the parameter $\lambda_3$. The Hamiltonian $\ham_{0,0}$ is known to be integrable in the sense of Bethe ansatz and to possess an extensive number of LIOMs~\cite{tovmasyanPreformedPairsFlat2018}. The results regarding level spacing statistics make a compelling case that the same also holds true for the more general OBS Hamiltonian $\ham_{0,\lambda_3}$.
 
The next natural step would be trying to construct the LIOMs directly. Whereas generally difficult, the task should be somewhat simpler in the case of the OBS model than for the dice lattice. There are at least two different strategies that can be followed. To begin, one could try some sort of perturbative expansion in the small parameter $\lambda_3$, since for the unperturbed case the LIOMs take a particularly simple form. Another option is to diagonalize the OBS model using the Bethe anstaz method. The study of the  scattering problem in Sec.~\ref{sec:few_body_problem} is a first step in this direction. Hamiltonians integrable in the Bethe Ansatz sense are rare and hard to come by, but at the same time have provided essential insights into the complexity and richness of quantum many-body systems. Showing that the OBS model belongs to this restricted class would be an invaluable and potentially useful result  for understanding flat band superconductors. However, one should note that there are several different notions of integrability in the quantum context and a Poissonian, or close to Poissonian, level spacing distribution does not automatically implies integrability in the Bethe ansatz sense~\cite{cauxRemarksNotionQuantum2011}.

Finally, it would be interesting to also extend the level spacing statistics analysis presented here to lattice models with flat bands in two dimensions. The most natural candidates are  the dice lattice  or the kagome lattice, which has been realized with ultracold gases in optical lattices \cite{joUltracoldAtomsTunable2012, leungInteractionEnhancedGroupVelocity2020, barterSpatialCoherenceStrongly2020}. The two-dimensional case is more difficult due to the absence of an integrable limit in which the Hamiltonian is expected to have a Poissonian level spacing distribution and to the additional spatial symmetries. Nevertheless, the experience gained with the OBS model will be certainly invaluable and our hope is that the analysis of level spacing statistics will lead to a deeper understanding of the nature of excitations in flat band superconductors.

\begin{acknowledgments}
We thank Jos\'e Lado and P\"aivi T\"orm\"a for useful discussions and their helpful comments on this paper. We acknowledge support from the Research Council of Finland under Grants No. 330384, No. 336369, and No. 358150. We acknowledge the computational resources provided by the Aalto Science-IT project. 
\end{acknowledgments}

\bibliography{references}

\end{document}